\documentclass[fleqn,usenatbib]{mnras}

\usepackage{newtxtext,newtxmath}

\usepackage[T1]{fontenc}
\usepackage{ae,aecompl}

\usepackage{graphicx}	
\usepackage{amsmath}	
\usepackage{amssymb}	
\usepackage{gensymb}    
\usepackage{upgreek}

\title[A radio pulsar with a 12.1 second period]{The SUrvey for Pulsars and Extragalactic Radio Bursts IV: Discovery and polarimetry of a 12.1-second radio pulsar}

\author[V. Morello et al.]{V. Morello,$^{1}$\thanks{E-mail: vincent.morello@postgrad.manchester.ac.uk}
E.F. Keane,$^{2}$
T. Enoto,$^{3}$
S. Guillot,$^{4}$
W.C.G. Ho,$^{5}$
A. Jameson,$^{7,16}$
\newauthor
M. Kramer,$^{6,1}$
B.W. Stappers,$^{1}$
M. Bailes,$^{7,16}$
E.D. Barr,$^{6}$
S. Bhandari,$^{8}$
M. Caleb,$^{1}$
\newauthor
C.M.L. Flynn,$^{7}$
F. Jankowski,$^{1}$
S. Johnston,$^{8}$
W. van Straten,$^{9}$
Z. Arzoumanian,$^{10}$
\newauthor
S. Bogdanov,$^{11}$
K.C. Gendreau,$^{10}$
C. Malacaria,$^{12,15}$
P.S. Ray,$^{13}$
R.A. Remillard$^{14}$
\\
$^{1}$Jodrell Bank Centre for Astrophysics, School of Physics and Astronomy, The University of Manchester, M13 9PL, UK\\
$^{2}$SKA Organisation, Jodrell Bank Observatory, SK11 9DL, UK\\
$^{3}$Department of Astronomy, Kyoto University, Kitashirakawa-Oiwake-cho, Sakyo-ku, Kyoto 606-8502,  Japan\\
$^{4}$IRAP, CNRS, 9 avenue du Colonel Roche, BP 44346, F-31028 Toulouse Cedex 4, France\\
$^{5}$Department of Physics and Astronomy, Haverford College, 370 Lancaster Avenue, Haverford, PA 19041, USA\\
$^{6}$Max-Planck-Institut f\"ur Radioastronomie, Auf dem H\"ugel 69, D-53121 Bonn, Germany\\
$^{7}$Centre for Astrophysics and Supercomputing, Swinburne University of Technology, PO Box 218, Hawthorn, VIC 3122, Australia\\
$^{8}$CSIRO Astronomy and Space Science, PO Box 76, Epping, NSW 1710, Australia\\
$^{9}$Institute for Radio Astronomy \& Space Research, Auckland University of Technology, Private Bag 92006, Auckland 1142, New Zealand\\
$^{10}$X-ray Astrophysics Laboratory, Astrophysics Science Division, NASA's Goddard Space Flight Center, Greenbelt, MD 20771, USA\\
$^{11}$Columbia Astrophysics Laboratory, Columbia University, 550 West 120th Street, New York, NY, 10027, USA\\
$^{12}$NASA Marshall Space Flight Center, NSSTC, 320 Sparkman Drive, Huntsville, AL 35805, USA\\
$^{13}$Space Science Division, U.S. Naval Research Laboratory, Washington, DC 20375 USA\\
$^{14}$MIT Kavli Institute for Astrophysics and Space Research, 70 Vassar Street, Cambridge, MA 02139, USA\\
$^{15}$Universities Space Research Association, NSSTC, 320 Sparkman Drive, Huntsville, AL 35805, USA\\
$^{16}$OzGrav - ARC Centre of Excellence for Gravitational Wave Discovery, PO Box 218, Hawthorn, Vic, Australia
}

\date{Accepted XXX. Received YYY; in original form ZZZ}

\pubyear{2019}

\begin{document}
\label{firstpage}
\pagerange{\pageref{firstpage}--\pageref{lastpage}}
\maketitle

\begin{abstract}
We report the discovery of PSR~J2251$-$3711, a radio pulsar with a spin period of 12.1 seconds, the second longest currently known. Its timing parameters imply a characteristic age of 15 Myr, a surface magnetic field of $1.3 \times 10^{13}$~G and a spin-down luminosity of $2.9 \times 10^{29}~\mathrm{erg~s}^{-1}$. Its dispersion measure of 12.12(1)~$\mathrm{pc}~\mathrm{cm}^{-3}$ leads to distance estimates of 0.5 and 1.3 kpc according to the NE2001 and YMW16 Galactic free electron density models, respectively. Some of its single pulses show an uninterrupted 180 degree sweep of the phase-resolved polarization position angle, with an S-shape reminiscent of the rotating vector model prediction. However, the fact that this sweep occurs at different phases from one pulse to another is remarkable and without straightforward explanation. Although PSR~J2251$-$3711 lies in the region of the $P-\dot{P}$ parameter space occupied by the X-ray Isolated Neutron Stars (XINS), there is no evidence for an X-ray counterpart in our \textit{Swift} XRT observation; this places a 99\%-confidence upper bound on its unabsorbed bolometric thermal luminosity of $1.1 \times 10^{31}~(d / 1~\mathrm{kpc})^2~\mathrm{erg/s}$ for an assumed temperature of 85 eV, where $d$ is the distance to the pulsar. Further observations are needed to determine whether it is a rotation-powered pulsar with a true age of at least several Myr, or a much younger object such as an XINS or a recently cooled magnetar. Extreme specimens like PSR J2251$-$3711 help bridge populations in the so-called neutron star zoo in an attempt to understand their origins and evolution.
\end{abstract}



\section{Introduction}
The standard evolutionary picture of isolated pulsars is that they are born with spin periods of up to a few tens of milliseconds~\citep{Noutsos2013,Lyne2015} and, through the loss of rotational energy via electromagnetic radiation and other processes, slow down~\citep[see e.g.,][]{lgs12}. The star cools quickly, by an order of magnitude in $\lesssim 10^6$~yr ~\citep[e.g.,][]{ppp15}, such that older neutron stars are generally difficult or impossible to detect via their thermal X-ray emission. Those that produce radio emission do so for $\sim 10^{7}$~yr before the emission mechanism quenches~\citep{kmk+13}. The exact criteria for cessation of radio emission are not well constrained but are generally considered to be dependent upon the magnetic field strength at the stellar surface and the spin period \citep{cr93}. The various models predict a lack of radio emission in the so-called `death valley' region of the spin period--period derivative ($P-\dot{P}$) parameter space where there is a dearth of sources. Although not accounted for by models, there is mounting observational evidence that the cessation does not happen abruptly, with intermittent emission observed now in several sources~\citep{mlo+06,crc+12,yws+14}.

Radio pulsars have been found with a wide range of spin periods, stretching from $1.4$~ms to $23.5$~s~\citep{hrs+06,tbc+18}, such that there now appears to be a range of spin periods, in the tens of seconds, that is occupied by both radio emitting neutron stars and white dwarfs~\citep{pat79}. The millisecond pulsars are understood to have been `spun up' through interactions with their later-evolving binary partners \citep{bh91}. This spin-up process reignites the radio emission process if it has ended; in contrast, isolated pulsars remain dead. It is therefore with the isolated pulsars that one could trace out the true spin evolution post-supernova if their ages could be accurately determined; unfortunately, that is precisely where the difficulty lies. Accurate ages have been obtained via supernova remnant (SNR) associations \citep{gf00, klh+03}, or by tracing pulsar trajectories in the Galactic gravitational potential back to their most likely birth site(s), yielding a so-called \textit{kinematic age} \citep[e.g., ][]{Noutsos2013}. These methods are not always applicable; SNRs remain detectable only for $\simeq 10^5$ yr, while the derivation of a kinematic age requires at least a proper motion measurement. In the absence of a better alternative, as is commonly the case with older ($\gtrsim 1$ Myr) neutron stars, one has to fall back to the inaccurate but available `characteristic age' $\tau_c = P/(2\dot{P})$.

Radio pulsars are not the only representatives of isolated neutron stars (INS), which manifest themselves under several sub-types of objects with different observational properties. Two classes are of particular interest here. Magnetars derive their name from the high surface magnetic fields ($B_{\mathrm{surf}} \simeq 10^{13} - 10^{15}$ G) implied by their spin-down parameters. They are typically very young ($\simeq 1 - 30$ kyr) and have spin periods of a few seconds. Their X-ray and soft $\gamma$-ray emission is generally accepted to be powered by the dissipation of their intense magnetic fields \citep{td95, kb17}, since their average luminosity often greatly exceeds their rotational energy loss rate. X-ray Isolated Neutron Stars (XINS) are nearby cooling neutron stars characterized by their thermal emission in the soft X-ray band and overall steadiness as sources. Their spin periods lie in the $3 - 17$\,s range and their ages are estimated to be of a few hundred kyr \citep{Turolla2009}. XINS have never been seen in the radio domain despite extensive attempts, although this could just be caused by a chance misalignment of their hypothetical radio beams with our line of sight \citep{Kondratiev2009}. The fact that the radio pulsar PSR J0726$-$2612 was found to produce thermal emission similar to that of the XINS \citep{rms+19} supports the misalignment hypothesis. For a complete overview of the INS diversity, one can refer to a review paper such as \citet{KaspiKramer2016}.

Many of the INS class labels are not mutually exclusive and, most importantly, it has been shown that the birth rate of core-collapse supernovae is insufficient to account for all neutron stars being born directly into whichever sub-category they currently belong to; this implies that there must be evolutionary links between at least some classes of INS \citep{KeaneKramer2008}. Correctly identifying all such possible links is key to achieving what has been previously called the unification of the neutron star zoo \citep{Kaspi2010}. If this is the goal, then it is worth trying to uncover the possible ancestors of the newly emerging population of very slow-spinning radio pulsars.

In this paper we describe, in \S\ref{sec:radio_observations}, the discovery and subsequent radio observations of PSR~J2251$-$3711, a $12.1$-second period radio pulsar. In \S\ref{sec:radio_analysis} we provide its phase-coherent timing solution and perform detailed analysis of its radio emission characteristics. \S\ref{sec:xray} describes the observations of the source in the X-rays with the \textit{Neil Gehrels Swift X-ray Observatory} \citep{Burrows2005} and the \textit{Neutron Star Interior Composition Explorer (NICER)}. In \S\ref{sec:disc} we attempt to constrain the true age of this pulsar and discuss where it may fit in an evolutionary context with respect to the entire population, before drawing our conclusions in \S\ref{sec:conc}.

\section{Radio Observations}
\label{sec:radio_observations}

\subsection{Discovery}

\begin{figure*}
\includegraphics[width=0.85\textwidth]{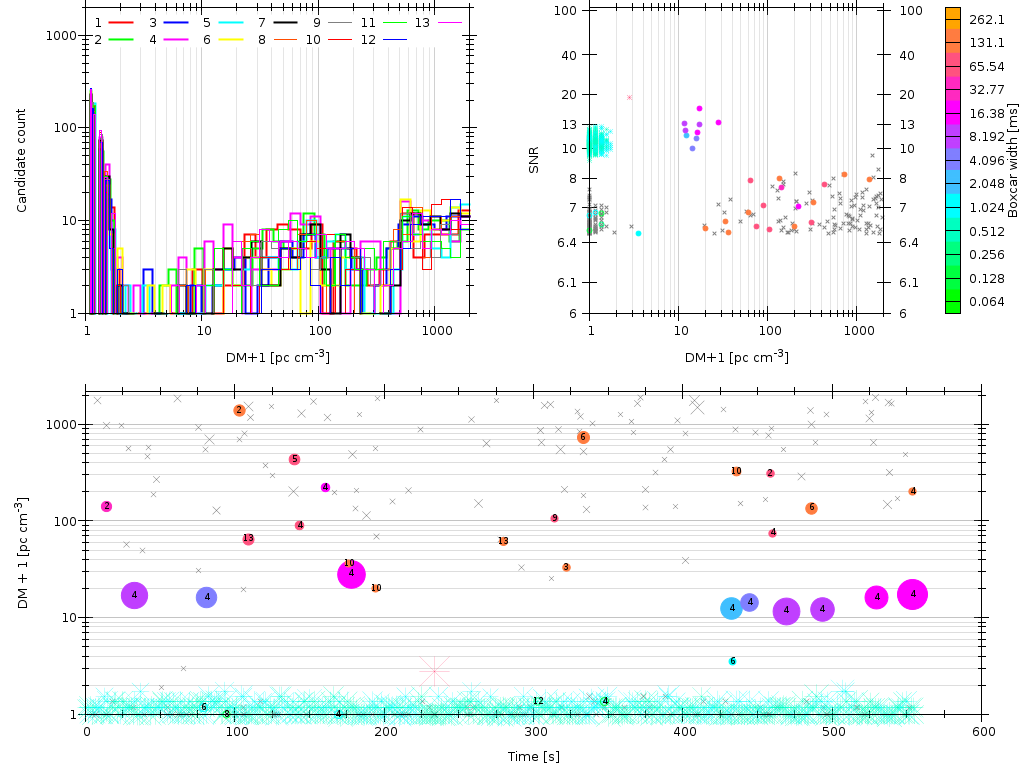}
\caption{Discovery plot of PSR~J2251$-$3711 as produced by the \textsc{heimdall} single pulse search pipeline. Detected pulses in this 9-minute discovery observation are shown as filled circles in the bottom panel in a time - dispersion measure (DM) diagram. The colour denotes the value of the best-fit pulse width, and the circle size indicates signal-to-noise ratio (SNR). The beam number in which they were detected is overlaid. Top left: DM distributions of detected single pulses, plotted for every individual beam. Top right: detected single pulses in a DM-SNR diagram, for all 13 beams of the receiver.}
\label{fig:discovery_plot}
\end{figure*}

PSR~J2251$-$3711 was discovered in the SUrvey for Pulsars and Extragalactic Radio Bursts ~\citep[SUPERB, see][for details]{SUPERBI}, conducted with the Parkes 21-cm multibeam receiver \citep{ParkesMultibeamReceiver}. At the time of discovery, SUPERB was using a Fourier-domain search as well as a single pulse search and it was in the latter that the pulsar was first detected, in a 9-minute blind survey observation taken on 8 Dec 2015 (UTC 2015-12-08-10:06:21, beam number 4). Figure \ref{fig:discovery_plot} shows the single pulse search diagnostic plots, exhibiting 9 pulses detected with signal-to-noise ratios in excess of 10$\sigma$ and best-fit widths between 4 and 16~ms. A rudimentary analysis of the differences between pulse times of arrival (TOAs) in this 9-minute discovery observation alone initially suggested a best-fit period of 6.06 s; however, no signal was detected by directly folding the raw data. The true period of 12.12 s was serendipitously found in January 2016 while testing a Fast Folding Algorithm (FFA) search code (then in an early phase of development) in conjunction with a multi-beam interference mitigation code. In a 1-hour confirmation observation taken on 19 Jan 2016, the source was seen again with a much higher statistical significance and any remaining doubts on the true period were dissipated.

We also looked for archival observations near the position of PSR~J2251$-$3711, and found that two dispersed pulses from the source were visible in a 4.5-minute observation in the high Galactic latitude portion of the High Time Resolution Universe \citep[HTRU, ][]{kjs+10} survey (UTC 2009-05-02-01:08:57, beam number 7). The source likely went undetected in the HTRU single pulse search due to its low dispersion measure ($\mathrm{DM} = 12.1~ \mathrm{pc~cm}^{-3}$, see \S \ref{subsec:dm}) making it difficult to distinguish from radio-frequency interference (RFI) that occurs predominantly at DM~0 \citep{kle+10}.

\subsection{Parkes multibeam receiver observations}

In order to obtain a phase-coherent timing solution, PSR~J2251$-$3711 has been observed regularly since its discovery whenever the SUPERB project was allocated telescope time at Parkes, except for a 10-month hiatus between Feb and Dec 2016 during which the multibeam receiver was taken down for refurbishment. All timing observations were acquired with the Berkeley Parkes Swinburne Recorder (BPSR) backend, using the standard search-mode SUPERB configuration: a 1382 MHz centre frequency, 400 MHz of bandwidth divided in 1024 frequency channels and a time resolution of 64 $\upmu$s. Only Stokes I was recorded with 2-bit digitization and integration times were either 18 or 30 minutes. In addition to the regular timing campaign, a longer 2-hour observation was taken on 28 Sep 2018, where all four Stokes parameters were recorded with the BPSR backend as well, this time with 8-bit precision but with the number of frequency channels reduced to 128 due to system limitations. From this observation, henceforth referred to as \textit{the main observation}, we were able to obtain polarization profiles and perform a number of single pulse analyses.

\section{Radio Analysis}
\label{sec:radio_analysis}

\subsection{Dispersion measure and distance estimation}
\label{subsec:dm}

The dispersion measure of a pulsar is routinely used to estimate its distance using a model of the Galactic free electron density, the two most recent and widely used being NE2001 \citep{NE2001} and YMW16 \citep{ymw16}. While in most cases the uncertainty on the distance thus derived is dominated by model limitations, in the case of PSR~J2251$-$3711 the dispersion delay between edges of the Parkes multibeam receiver band is approximately half its integrated pulse width. This corresponds to a large fractional error on the DM  which therefore contributes significant additional uncertainty to a distance estimate. Indeed, running a period-DM optimization utility such as \textsc{psrchive}'s \texttt{pdmp} \citep{PSRCHIVE} on the folded main observation yields a best fit DM of $15.8 \pm 3.6~\mathrm{pc~cm}^{-3}$. However, we can take advantage of two facts: first, that individual single pulses from PSR~J2251$-$3711 are considerably narrower than the integrated pulse (see \S \ref{subsec:single_pulse_intensity}), and second, that each of them acts as an \textit{independent} DM estimator. Therefore, a large number of detectable single pulses can in principle be combined into a highly accurate DM estimator, within the 11 hours worth of data in which the source is visible.
 
We therefore ran the \textsc{heimdall}\footnote{\url{https://sourceforge.net/projects/heimdall-astro/}} single pulse search pipeline on all Parkes observations of PSR~J2251$-$3711 with a narrow DM search step of 0.04 $\mathrm{pc~cm}^{-3}$. This returned a list of detected pulse DMs, signal-to-noise ratios (SNRs) and widths; the best reported width is that of the boxcar matched filter yielding the optimum response, noting that trial boxcar widths are equal to $2^k \times \tau$, where $k$ is an integer in the range [1, 12] and $\tau = 64~\upmu$s the time resolution of the data. After filtering out statistically insignificant events (SNR < 10), the remaining pulses showed a clearly bi-modal distribution in DM, with RFI events clustering around zero and pulses from the source around $\sim 12~\mathrm{pc~cm}^{-3}$. To completely eliminate any overlap between these two clusters, we also removed from the sample any pulse with a reported width W > 10 ms, which should not exclude many events originating from PSR~J2251$-$3711. Finally, we removed all zero-DM events and a total of 326 pulses were left in our sample.

One missing ingredient here is the set of uncertainties on the reported single pulse DMs, but they can be inferred from the data as described below. We assumed them to be Gaussian with standard deviations $\sigma_i$ proportional to the pulse widths $w_i$, which we choose to express as
\begin{equation}
\label{eq:dm_uncertainty}
\sigma_i = f \times \frac{w_i}{k_{\mathrm{DM}} \left( \nu_{\mathrm{min}}^{-2} - \nu_{\mathrm{max}}^{-2} \right)}
\end{equation}
where $k_{\mathrm{DM}} = 4.148808 \times 10^3~\mathrm{pc}^{-1}~\mathrm{cm}^3~\mathrm{MHz}^2~\mathrm{s}$ is the dispersion constant, $\nu_{\mathrm{min}} = 1182$ MHz and $\nu_{\mathrm{max}} = 1523.4$ MHz are the bottom and top effective\footnote{the top 58.6 MHz of the band are almost permanently occupied by RFI and were masked in the search} observing frequencies expressed in MHz, $w_i$ the reported pulse width in seconds, and $f$ is a dimensionless, \textit{a priori} unknown uncertainty scale factor to be fitted along other model parameters. In clearer terms, we have written that $\sigma_i$ is proportional to the DM that corresponds to a dispersion delay $w_i$ across the observing band.

\begin{figure}
    \centering
    \includegraphics[width=1.00\columnwidth]{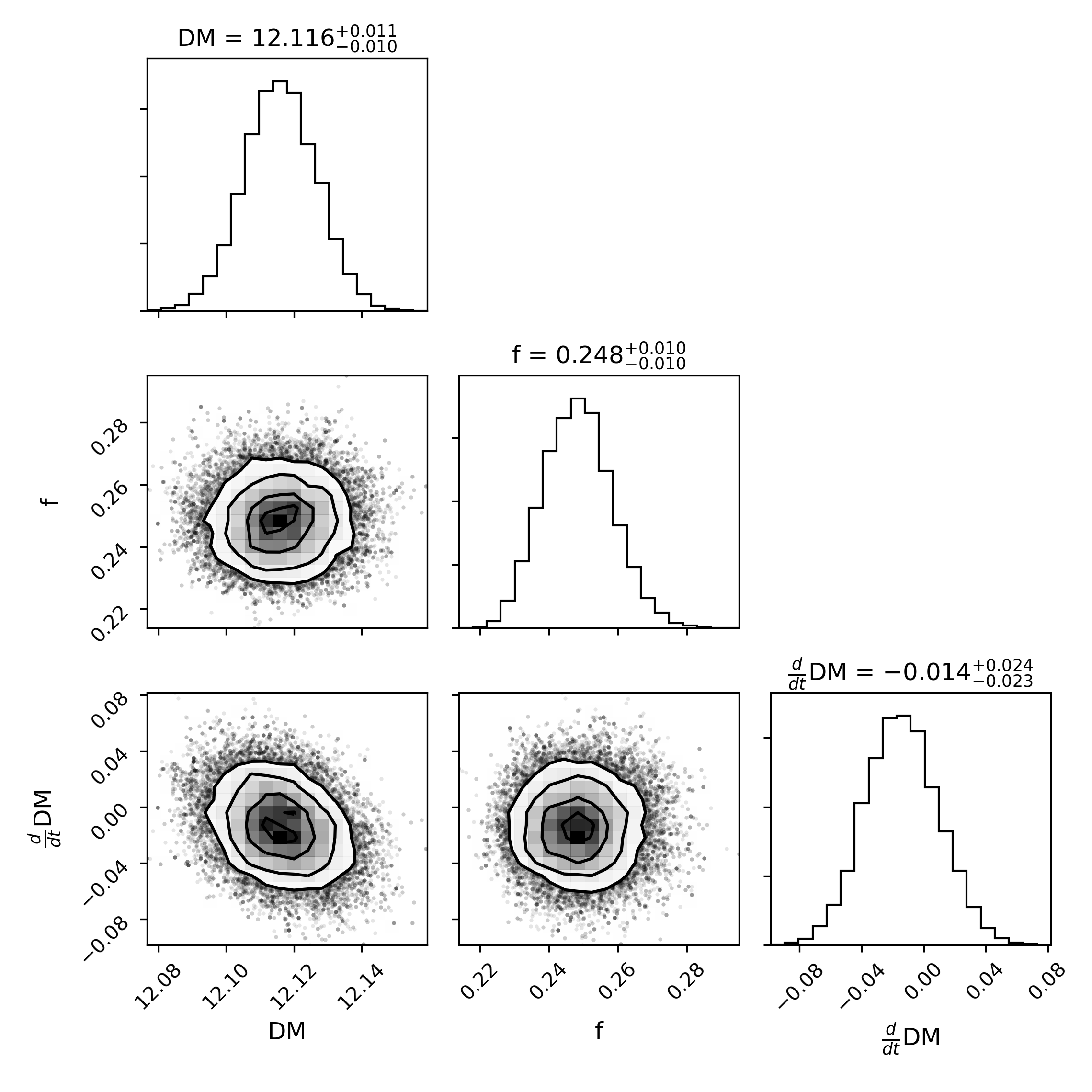}
    \caption{Posterior probability densities for the source dispersion measure, dispersion measure rate of change (here in $\mathrm{pc~cm}^{-3}\mathrm{yr}^{-1}$), and the dimensionless DM uncertainty scale factor $f$ (see Eq. \ref{eq:dm_uncertainty} and text for details). They were fitted to a sample of 326 single pulses detected by the search pipeline \textsc{heimdall} across all Parkes multibeam observations of PSR~J2251$-$3711. The lower and upper uncertainties quoted for all parameters correspond respectively to the 16th and 84th percentiles of their probability distributions.}
    \label{fig:posterior_distributions}
\end{figure}

Given the set of $n = 326$ observations $(x_i, w_i, t_i)$, denoting respectively the observed DM, width and MJD of detection of every pulse, the associated log-likelihood function can be written

\begin{equation}
\label{eq:log_likelihood_dmfit}
\ln \mathcal{L}(d, \dot{d}, f) = - \sum_{i=1}^{n} \frac{\left(x_i - d - \dot{d}(t_i - t_{\mathrm{ref}})\right)^2}{2 \sigma_i^2} -\frac{1}{2} \sum_{i=1}^{n} \ln{\left( 2 \pi \sigma_i^2 \right)},
\end{equation}
where $d$ is the source DM, $\dot{d}$ the secular DM variation rate in $\mathrm{pc~cm}^{-3}\,\mathrm{day}^{-1}$ and $t_{\mathrm{ref}} = 57900$ the reference MJD of the fit. The first term is essentially a reduced chi-square, and the second term places a penalty on higher values of $f$. We used the Markov chain Monte Carlo ensemble sampling package \texttt{emcee} \citep{emcee} to estimate the joint posterior probability distribution of $d$, $\dot{d}$ and $f$ (Figure \ref{fig:posterior_distributions}). We assumed uniform prior distributions for $\dot{d}$ and $f$, and a normally distributed prior for $d$ mean and standard deviation given by the best DM and DM uncertainty reported by running \texttt{pdmp} on the folded main observation. We obtained $d = 12.12 \pm 0.01~\mathrm{pc~cm}^{-3}$ with $f = 0.25 \pm 0.01$, and found no measurable secular DM variation (Figure \ref{fig:posterior_distributions}). We verified \textit{a posteriori} the model assumptions by inspecting the fit residuals $\left(x_i - d - \dot{d}(t_i - t_{\mathrm{ref}})\right) / \sigma_i$, which were found to have a median value of 0.03 and standard deviation of 1.0, i.e. consistent with the expected normal distribution.

From the DM value obtained, the Galactic electron density models NE2001 and YMW16 predict distances of 0.54~kpc and 1.3~kpc respectively. Considering that the distance to the pulsar is a key parameter when attempting to constrain its X-ray luminosity (see \S \ref{sec:xray}), this relative discrepancy is both large and unfortunate. On a sample of 189 known pulsars where a more reliable independent distance estimate is available, both NE2001 and YMW16 distance predictions were found to be inconsistent by a factor of 2 or more about 20\% of the time \citep[Figure 14 of][]{ymw16}. More recently, it has been shown from a sample of 57 pulsar parallaxes determined via very long baseline interferometry, that DM distances must be treated with even more caution for sources that are either nearby or at high Galactic latitudes \citep{dgb+19}. As PSR~J2251$-$3711 arguably belongs to both categories, we need to remain open to the possibility that its distance may lie outside of the [0.5, 1.3] kpc range.

\subsection{Timing analysis}

All available observations were dedispersed at the DM determined in the analysis above and folded using \texttt{dspsr} \citep{vanStraten2011}. We fitted a single von Mises\footnote{a function defined on the unit circle, with a shape similar to a Gaussian} component to the integrated pulse profile of the main observation using the \texttt{paas} utility of \textsc{psrchive}, and the resulting noise-free pulse template was used to extract pulse times of arrival (TOAs) from every folded observation. A total of 40 reliable TOAs were obtained, to which we fitted a phase-connected timing solution using \textsc{tempo2} \citep{TEMPO2}. We limited the fit parameters to position, spin frequency, and spin frequency derivative, with the dispersion measure parameter being excluded and fixed to the value of $\mathrm{DM} = 12.12~\mathrm{pc~cm}^{-3}$ previously obtained. The resulting timing solution is presented in Table \ref{tab:ephemeris}.

We note that the HTRU high-latitude survey detection of 2009 is significant enough to yield a TOA. It however deviates by approximately $\Delta t = 300$ ms from the time of arrival predicted by the ephemeris given in Table \ref{tab:ephemeris}. Since $\Delta t$ can in principle be further off an integer multiple of $P$, there is a non-negligible $2 |\Delta t|/ P \approx 5\%$ probability for the TOA to fall within |$\Delta t$| of the prediction by chance alone. We have therefore excluded it from this analysis. Including it in the fit yields $P$ and $\dot{P}$ values consistent with those reported in Table \ref{tab:ephemeris} with uncertainties 5 times smaller, but this does not improve the positional uncertainty and more importantly does not enable an accurate fit for extra parameters such as proper motion or a second period derivative.

\begin{figure}
\includegraphics[width=1.00\columnwidth]{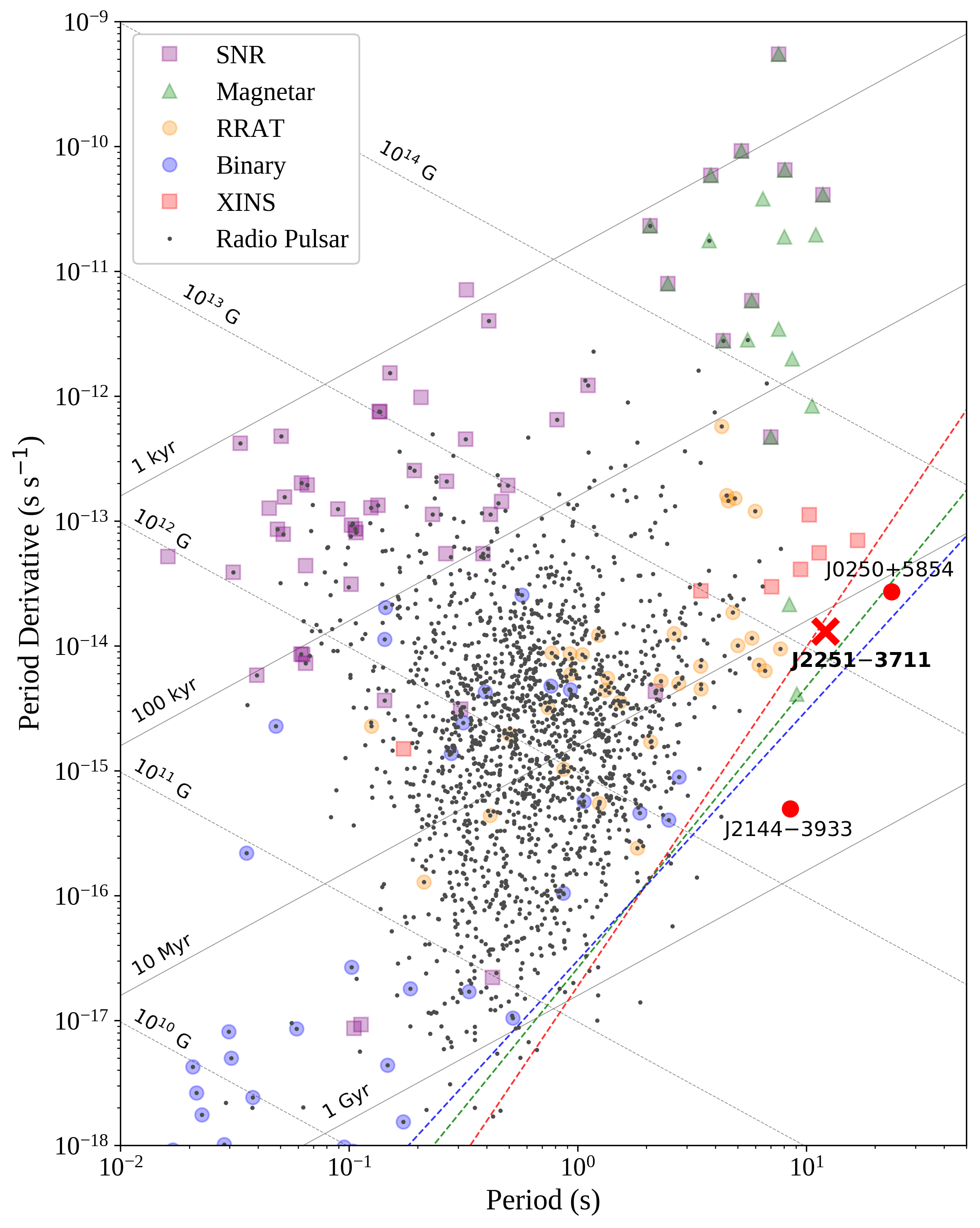}
\centering
\caption{$P-\dot{P}$ diagram, based on v1.59 of the ATNF pulsar catalogue \citep{PSRCAT}. The period and period derivative ranges on the plot have been set to concentrate on the non-recycled pulsar population. The radio pulsars with the three longest spin periods have been highlighted in red. Lines of constant characteristic age and surface magnetic field strength are displayed in grey. The dashed red line represents the lower limit of the so-called pulsar death valley \citep[Eq. 8 of][]{cr93}. Death lines from \citet{zhm00} are also shown, based on their curvature radiation from vacuum gap model (green dashes) and space-charged-limited flow model (blue dashes).}
\label{fig:ppdot}
\end{figure}

The location of PSR~J2251$-$3711 on the $P-\dot{P}$ diagram is presented in Figure \ref{fig:ppdot}, highlighted as a red cross. PSR~J2251$-$3711 lies very near a region in which radio emission is predicted to shut down according to classical emission models \citep{cr93, zhm00}; however, it does not challenge these models to the same extent as radio pulsars PSR~J0250$+$5854 \citep{tbc+18} and PSR~J2144$-$3933 \citep{ymj99}, which have spin periods of 23.5 seconds (longest known) and 8.5 seconds (third-longest known) respectively.

\begin{table}
\caption{PSR~J2251$-$3711 timing model and derived parameters. The dispersion measure was fitted on a large sample of single pulses (see \S \ref{subsec:dm} for details). The numbers in parentheses express the 1-$\sigma$ uncertainties on the last significant digit of each parameter.}
\label{tab:ephemeris}
\begin{tabular}{|l|l|}
  \hline
  \hline
  Timing Parameters                                  &  \\
  \hline
  Right Ascension, $\alpha$ (J2000)                  & 22:51:44.0(1) \\
  Declination, $\delta$ (J2000)                      & $-$37:11:48(2)  \\
  Spin Period, $P$ (s)                                & 12.122564931(1) \\
  Spin Period Derivative, $\dot{P}$ ($\rm{s~s^{-1}}$)       & $1.310(4) \times 10^{-14}$ \\
  Dispersion Measure, DM ($\rm{pc~cm^{-3}}$)         & 12.12(1) \\
  Spin Frequency, $\nu$ (Hz)                            & 0.082490793464(6) \\
  Spin Frequency Derivative, $\dot{\nu}$ ($\rm{Hz~s^{-1}}$)   & $-8.91(3) \times 10^{-17}$ \\
  Epoch of timing solution (MJD)                     & 57900 \\
  Timing span (MJD)                                  & 57363 $-$ 58598 \\
  Number of TOAs                                     & 40 \\
  RMS timing residual (ms)                           & 6.1 \\
  Solar system ephemeris model                       & DE430 \\
  Clock correction procedure                         & TT(TAI) \\
  \hline
  & \\
  Derived Parameters                                 &  \\
  \hline
  Galactic Longitude, $l$ (deg)                      & 3.603 \\
  Galactic Latitude, $b$ (deg)                       & $-$62.882 \\
  NE2001 DM Distance (kpc)                           & 0.54 \\
  YMW16 DM Distance (kpc)                            & 1.3 \\
  Characteristic Age (Myr)                           & 14.7 \\
  Characteristic Surface Magnetic Field (G)          & $1.3 \times 10^{13}$ \\
  Spin-down Luminosity (erg $\rm{s}^{-1}$)           & $2.9 \times 10^{29}$ \\
  \hline
\end{tabular}
\end{table}

\subsection{Long-term nulling}

Within a single observation, the emission of PSR~J2251$-$3711 is clearly sporadic, with more than half of its rotations showing no detectable pulse (\S \ref{subsec:single_pulse_intensity}). We therefore examined the entire SUPERB observation history to make an accurate census of non-detections of PSR~J2251$-$3711, in order to determine if these could be attributed to its emission actually shutting down for an extended period of time. Among a total of 57 radio observations taken since its discovery, the source was bright enough to yield a valid TOA in 40 of them. We phase-coherently folded the remaining 17 using the ephemeris in Table \ref{tab:ephemeris} and examined the resulting output for the presence of pulses. We confirmed the presence of statistically significant pulses from the source in all but 2 observations, which were respectively 35 and 17 minutes long. We note however that in both cases the RFI environment was particularly adverse, enough that it was impossible to reliably determine the actual source of a pulse occurring within the phase window expected to be occupied by the pulsar. As a result, we cannot entirely rule out that PSR~J2251$-$3711 would have been detected on both days had the observing conditions been quieter. Furthermore, and most importantly, the pulsar was successfully observed on the day following each of these non-detections. With the available data, we therefore find no compelling reason to believe that the radio emission from PSR~J2251$-$3711 could cease for several hours or longer.

\subsection{Mean polarization profiles}


We used the 2-hour long main observation to obtain the mean polarization profiles of the pulsar at 1382 MHz. Since the position angle (PA) of the linearly polarized flux is affected by Faraday rotation when propagating through the interstellar medium, it is necessary to first evaluate the rotation measure (RM) of the pulsar in order to determine the intrinsic PA at the pulsar. We ran the \texttt{rmfit} utility of \textsc{psrchive} on every single pulse, thus obtaining a set of measured pulse RMs $x_i$, RM uncertainties $\sigma_i$, and signal-to-noise ratios (SNRs). We filtered out statistically insignificant pulses and obvious outliers from the dataset, and then inspected the remaining pulses to ensure that they were originating from the pulsar and not from an interference source. In the end, we were left with $n=131$ reliable ($x_i$, $\sigma_i$) measurements. To determine the best-fit RM of the source, we used a similar method to that used in determining the DM (\S \ref{subsec:dm}). Our two fit parameters were the true rotation measure of the pulsar $r$ and a dimensionless uncertainty scale factor $f$, introduced to take into account any potential systematic under- or over-estimation of the $\sigma_i$ by the \texttt{rmfit} program. Ignoring constant terms, the log-likelihood of the dataset is

\begin{equation}
\label{eq:log_likelihood_rmfit}
\ln \mathcal{L}(r, f) = -n \ln f - \sum_{i=1}^{n} \frac{\left(x_i - r\right)^2}{2 f^2 \sigma_i^2},
\end{equation}
where we have postulated that the uncertainties on each pulse RM are normally distributed. Assuming uniform priors for $r$ and $f$ we obtained a best-fit RM value of $r = 11 \pm 1~\mathrm{rad~m}^{-2}$ and found that \texttt{rmfit} underestimated the uncertainties on every pulse RM by a factor $f = 1.6 \pm 0.1$. The fit residuals $(x_i - r) / (f \sigma_i)$ were consistent with the assumed normal distribution.

\begin{figure}
\includegraphics[width=1.00\columnwidth]{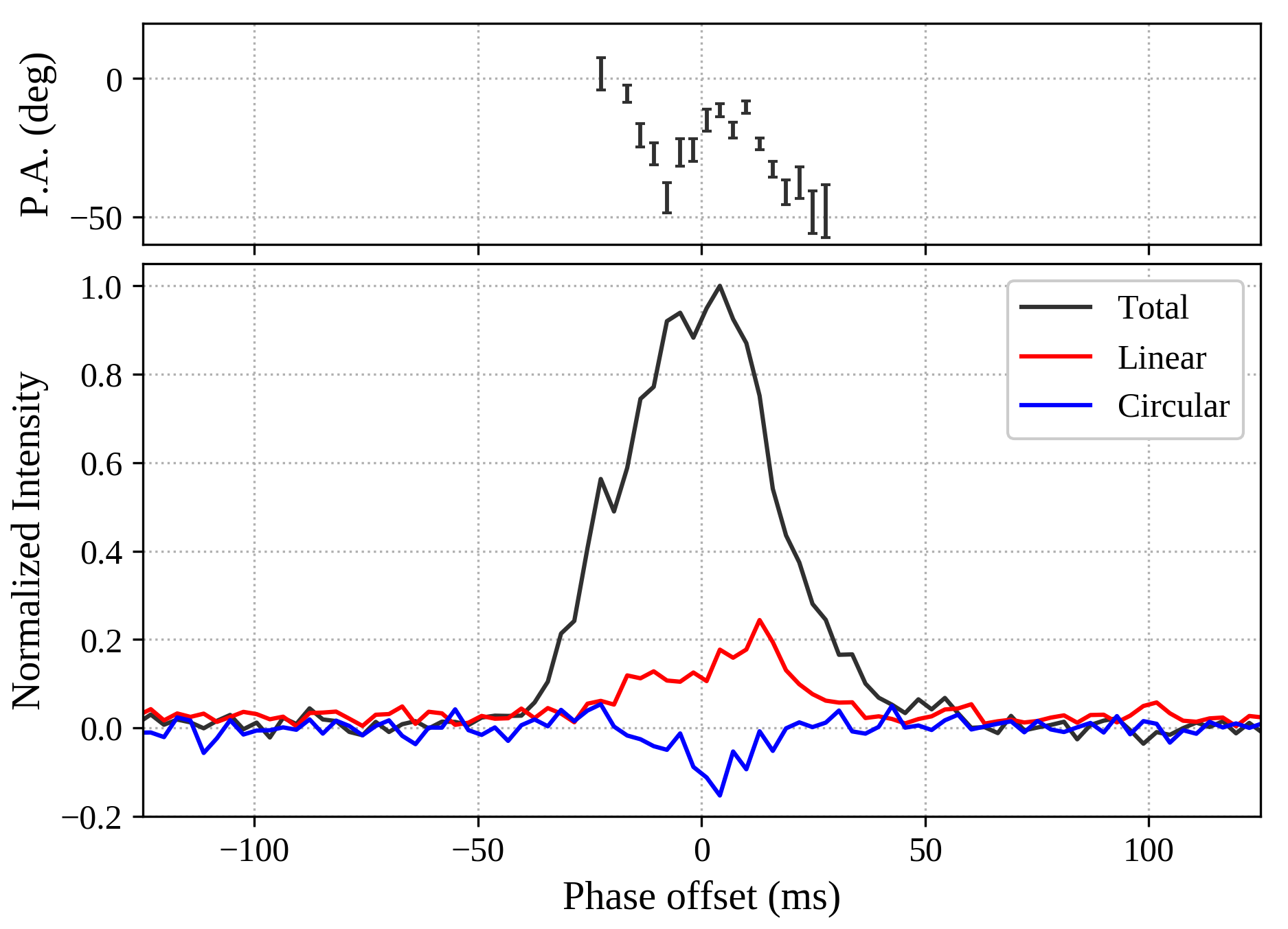}
\centering
\caption{Bottom panel: mean polarization profiles at 1382 MHz, normalized to peak total intensity and corrected for Faraday rotation. Top panel: position angle of the linearly polarized flux. Phase offset is measured from the peak of the Gaussian fit to the mean pulse intensity.}
\label{fig:polarization_profiles}
\end{figure}

The mean polarization profiles of PSR~J2251$-$3711 corrected for Faraday rotation are presented in Figure \ref{fig:polarization_profiles}, along with the position angle of the polarized flux. The fractions of linear and circular polarization are 17\% and 6\% respectively within the pulse phase region, defined as phase bins with total intensity at least 3 times larger than the standard deviation of the background noise. However, the mean profiles do not capture the more complex characteristics observed in single pulses, which show a much higher degree of polarization and interesting, if not puzzling behaviour of the PA (\S \ref{subsec:sppol}). Using the \texttt{scipy.optimize} python package, we found that the mean pulse intensity is well modeled by a single Gaussian component with a full width at half maximum (FWHM) $W_{50} = 40.0 \pm 0.3$ ms. The corresponding full width at ten percent of the maximum is $W_{10} = 72.9 \pm 0.5$ ms. The average flux density of the pulsar at 1.4 GHz, estimated from the radiometer equation on this specific observation, is $S_\mathrm{mean} = 0.15~\mathrm{mJy}$.

\subsection{Single pulse intensity analysis}
\label{subsec:single_pulse_intensity}

\begin{figure}
\includegraphics[width=0.85\columnwidth]{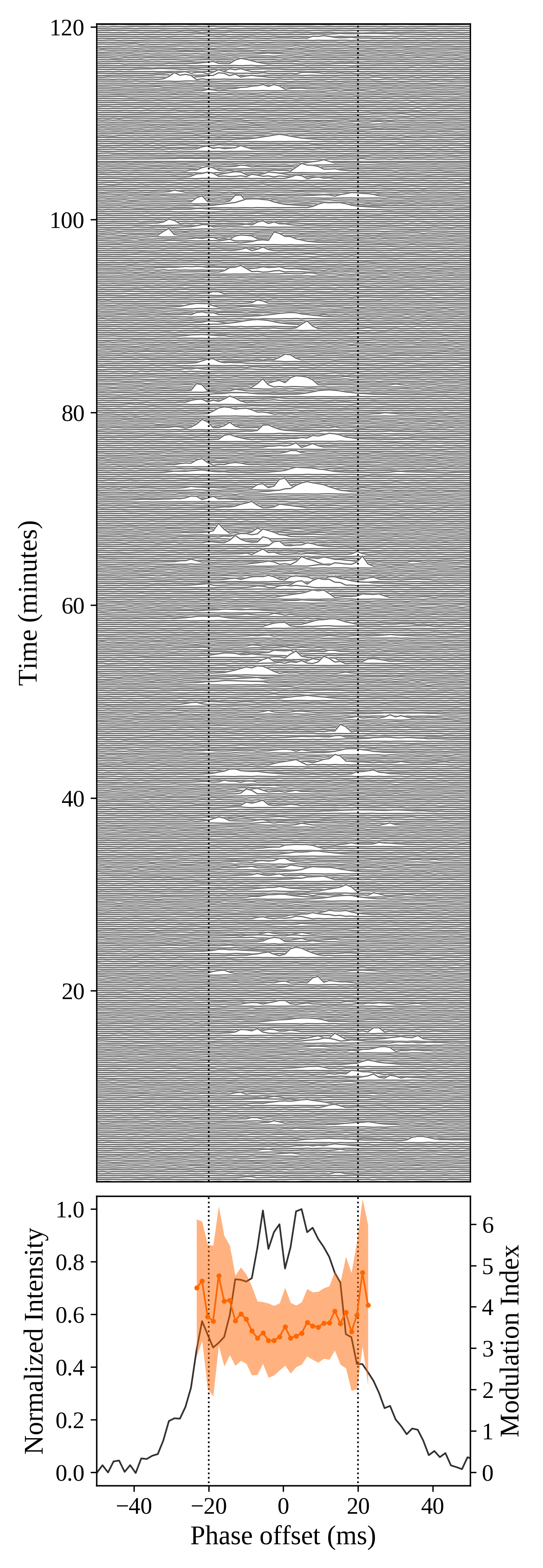}
\centering
\caption{Top panel: single pulses from a continuous 2h observation at
a centre frequency of 1382 MHz. The vertical lines denote the FWHM of the Gaussian pulse fit. Bottom panel: Phase-resolved flux density (grey line, left axis) and modulation index (orange points, right axis) around the on-pulse region. The 1-$\sigma$ errors are represented by the shaded area. The overall tendency of single pulses to arrive earlier as the observation progresses is statistically significant (see text), and could be evidence for the presence of drifting sub-pulses tracks, wrapping around in phase on a time scale of several hours.}
\label{fig:single_pulses}
\end{figure}

All single pulse intensities from the main observation are displayed in Figure \ref{fig:single_pulses}. At first glance, there is no clear evidence of sub-pulse drifting, as confirmed in a two-dimensional Fourier Transform of the single pulse stack. However, there was an indication of a systematic shift of pulse phase towards earlier time during this observation. A total drift of about 20ms can clearly be seen directly in Figure \ref{fig:single_pulses}. To verify this statement we measured the phase of every single pulse, taken to be the phase of the boxcar matched filter that gives the best response when convolved with the pulse. Fitting a straight line to the single pulse phases (in units of time) as a function of time confirms that the drift rate $r$ is statistically significant, with a value of $r = -3.6 \pm 0.6~ \mu\mathrm{s~s}^{-1}$. This find was an incentive to carefully double check whether the data had been folded at an incorrect period due to the pulsar ephemeris being wrong, but no issues were found, and no timing residual exceeds 20 ms. This phase drift can therefore not persist indefinitely, or has to be periodic with a peak-to-peak amplitude no larger than approximately one integrated pulse width $W_{50} = 40$ ms, otherwise it would manifest itself as a detectable periodic signal in the timing residuals. The best-fit total amount of drift in the main observation is $\Delta t = 26$ ms; if it was caused by an unmodeled movement of the source, it would correspond to a total line-of-sight (l.o.s.) displacement of $x = c \Delta t \simeq 8 \times 10^3$ km, which immediately rules out free precession of the pulsar as an explanation. The possibility of a binary companion then remains to be examined; if we assume that the pulsar follows a circular orbit, then the maximum l.o.s. orbital radius $R$ that could credibly remain undetectable in the timing data is $R = c W_{50} / 2$. Over the main observation, the pulsar would therefore have moved by approximately $x / 2R \simeq 65\%$ of an orbital diameter along the line of sight, implying that at least a quarter of the orbit has been covered, and therefore that the orbital period is no longer than 8 hours. Further assuming that the orbit is edge-on, solving Kepler's third law yields a minimum companion mass $m = 5 \times 10^{-3}~M_{\odot}$, about five jovian masses. However, the accepted formation scenario for such tight binary systems involves the pulsar spun-up to millisecond periods via accretion from a low-mass companion star, which appears quite unlikely here. The most reasonable explanation for the observed phase drift is therefore drifting sub-pulses over time scales longer than 2 hours. $r$ might also be an alias of a higher drift rate associated to one or more drifting sub-pulse tracks, but the intermittency of the emission makes this impossible to determine.

\begin{figure}
\includegraphics[width=1.00\columnwidth]{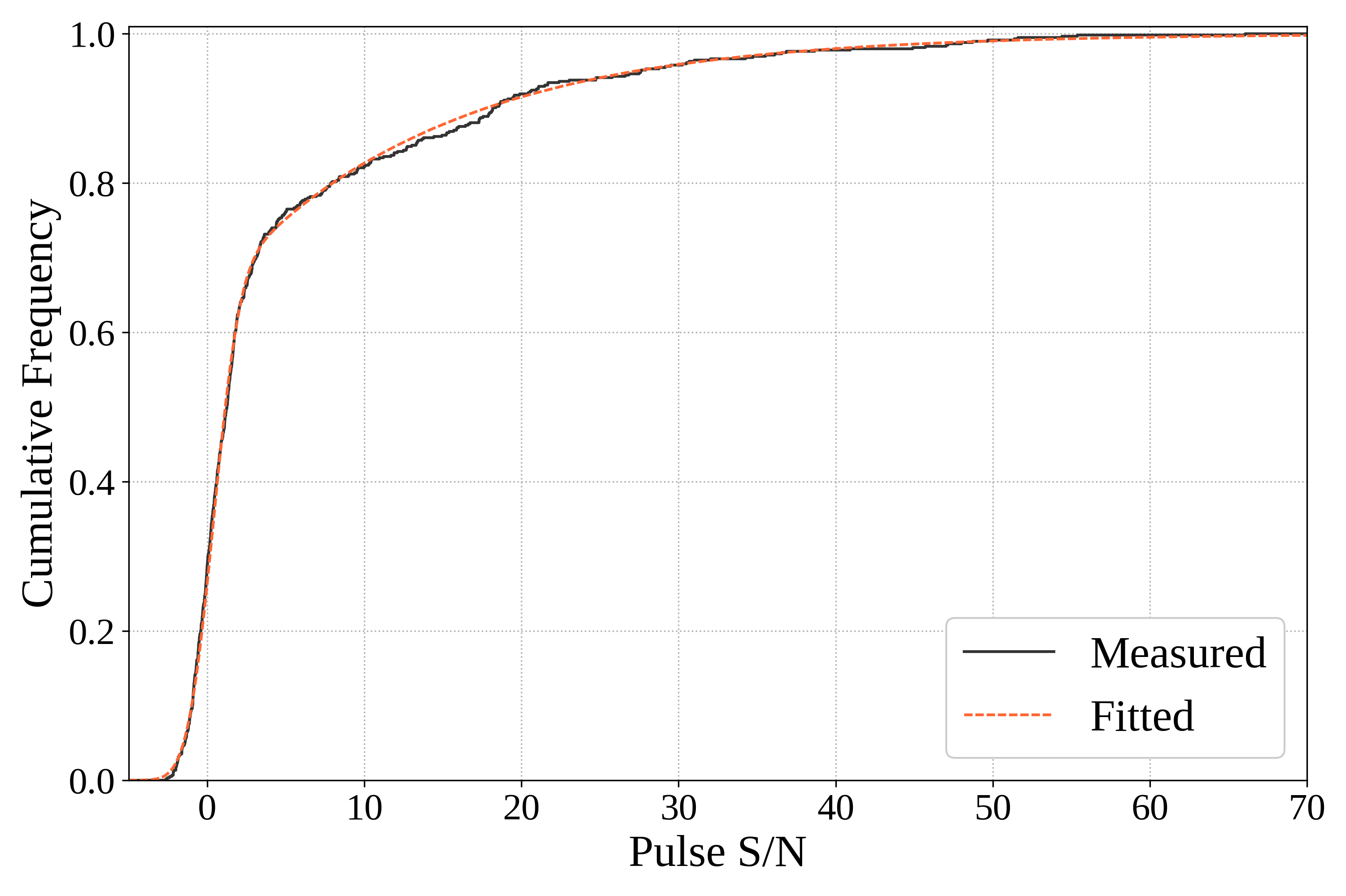}
\centering
\caption{Cumulative distribution function of single pulse energies measured on the 2-hour main observation (solid black line). S/N values are the integrated intensities of every pulse over an identical phase window, divided by the appropriate normalisation factor. The intrinsic pulse energy distribution is well fitted by an exponential distribution modulated by a nulling probability of 65\% (dashed orange line).}
\label{fig:pulse_energies}
\end{figure}

We also obtained a pulse energy distribution by measuring the pulse signal-to-noise ratios on an identical phase window for all pulses. The width of the window was chosen to be twice the FWHM of the Gaussian fit to the integrated pulse intensity $W_{50}$, to ensure that all the signal originating from the pulsar was accounted for. Using the \texttt{pdistFit} utility of the \textsc{psrsalsa} suite \citep{PSRSALSA}, we found that the pulse energy distribution is best fit by an exponential distribution with scale parameter $\lambda = 0.072$ modulated by a nulling probability of 65\%. The quantity $1/\lambda = 13.9$ represents the average S/N of pulses that are not nulls. Figure \ref{fig:pulse_energies} shows a comparison of both observed and fitted pulse energy cumulative distribution functions, which match closely.

\subsection{Single pulse polarization}
\label{subsec:sppol}

\begin{figure*}
\includegraphics[width=1.00\textwidth]{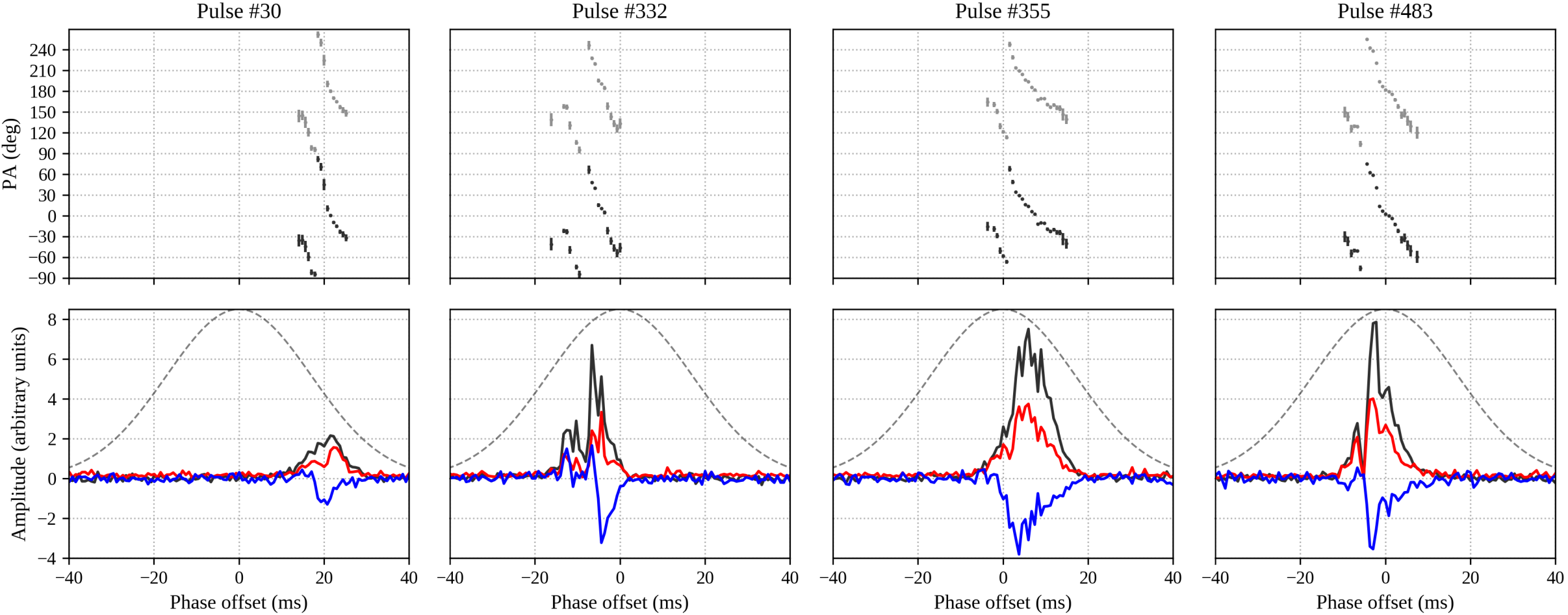}
\centering
\caption{Polarization profiles (at 1.4 GHz) of four example single pulses showing a smooth and uninterrupted 180 degree sweep of the position angle (PA) \textit{at different phases}. The time resolution of the data is 740 $\mathrm{\mu s}$. Top panel: phase-resolved PA (black points) where the data are repeated a second time with a shift of 180 degrees (grey points) for readability. Bottom panel: polarization profiles, with total intensity in black, linearly and circularly polarized flux in red and blue respectively. The grey dashed line represents the best-fit Gaussian to the integrated pulse intensity over the whole observation. Such continuous 180 degree rotation of the PA is somewhat reminiscent of the rotating vector model prediction for an inner line of sight (see text). However, the fact that the sweep occurs at different phases in every pulse is remarkable and difficult to explain. This behaviour is visible in a dozen pulses in the 2-hour observation.}
\label{fig:paswing180}
\end{figure*}

We examined the polarization profiles of all 596 single pulses, along with their phase-resolved position angle (PA) curves. Many PA curves are difficult to individually exploit, due to being incomplete as a result of low signal-to-noise ratio, insufficient linear polarization, or simply the absence of emission in some phase ranges. Still, at least 9 pulses show an uninterrupted 180-degree sweep of the PA with an S-shape similar to the rotating vector model \citep[RVM,][]{rc69} prediction. The unusual fact here is that the sweep occurs at significantly different phases from one pulse to the other. We have shown four examples of single pulses to illustrate this behaviour in Figure \ref{fig:paswing180}. Although some pulse periods show two or more well-spaced sub-pulses components, we note that there are no instances where two distinct 180-degree sweeps are observed. An RVM fit of the PA curve for these remarkable pulses does not provide any strong constraint on the emission beam geometry, due to a degeneracy between the angle between the spin and magnetic axes $\alpha$, and the impact parameter of the line of sight $\beta$. But we note that in all single pulses where a fully sampled 180-degree rotation of the PA is visible, we consistently observe two characteristics. Firstly, that the PA at the start of the sweep is close to zero degree. Secondly, that the PA monotonically decreases with phase; if one were to trust the RVM here, this would indicate that $\alpha > 90\degree$ and $\beta > 0\degree$, a so-called inner line of sight.

However, fully reconciling these observations with the RVM appears difficult. The main assumption of the RVM is that the direction of polarization is either parallel or orthogonal to the direction of the magnetic field at the point of emission; the field is taken to be strongly dipolar. A displacement of the point of emission along the plane orthogonal to the line of sight can explain a perceived delay (or advance) of the intensity curve, but should leave the PA curve invariant. It is then tempting to invoke changes of emission height between pulses, but the PA curve is expected to arrive $\Delta t = 4 r / c$ later than the intensity curve, where $r$ is the emission altitude \citep{bcw91}; clearly, both curves remain synchronised in all our pulses of interest, ruling out this explanation and requiring us to examine possible deviations from the RVM. In the radio pulsar population, the one most commonly observed is the so-called orthogonal polarization mode phenomenon \cite[OPM, e.g.,][]{mth75}, where the position angle of one radiation mode follows the main RVM swing while the other is offset by 90 degrees. Which mode is dominant can change as a function of pulse phase, causing so-called OPM transitions which register as 90-degree discontinuities in the position angle. While OPMs do seem to be present in PSR J2251$-$3711, transitions between modes can hardly account for a full 180 degrees worth of seemingly continuous PA rotation. If we were indeed to subtract 90 degrees worth of OPM transition from the steepest part of each PA curve in Fig. \ref{fig:paswing180}, we would still be left with another 90 degrees of residual smooth PA decrease that begins with an initial value of about 150 degrees and occurs at different phases, which still cannot be interpreted in light of the RVM. To fully account for the rotation solely with OPMs, we would have to be observing unusually smooth forward then backwards OPM transitions in immediate succession, where the second transition is perceived as a 90-degree rotation of PA in the same direction as the first (rather than opposite direction as one may expect). 


The time-shifted full-PA swings that are observed in the single pulses from PSR J2251$-$3711 may also be explainable by considering a multi-polar nature for the pulsar's magnetic field structure. One expects, far from the stellar surface, the dipolar contribution to dominate, whereas near the surface quadrupole or higher order moments contribute. It is usually considered to be the case that the field is dipolar in all locations of interest relevant to pulsar emission, but in this case, with a remarkably slow pulsar, the potentially very large magnetosphere (the light cylinder radius is $\sim 600,000$~km; for a 1-ms pulsar it would be $\sim 50$~km) means that the higher multipoles may play an appreciable role. Any deviations from pure dipolar behaviour very close to the stellar surface mean that multiple tangential emission beams, each pointed differently, could exist. For such a large magnetosphere the last closed field line encloses a small open field line region on the stellar surface such that if such a pulsar were observed at all then observing these multiple beams also might be more likely. Each individual beam would map out the same PA swing, but with relative time lags. We would expect to obtain the same $\beta$ and $\alpha$ values from RVM fits to each of the individual pulses, however as described above no constraining fit results when this is attempted. In this scenario we would also expect to see some instances of two (or more) full PA swings per pulse period; our sample does not contain examples of this.

With simple ideas failing to conclusively account for what is observed, more advanced explanations may have to be envisaged. For example, that during the time taken by the line of sight to cross the emission region of a given single pulse, the observer perceives significant temporal evolution of the physical properties of the emission region; or, that polarization of the radiation is \textit{not} determined at the point of emission, and is instead significantly altered by propagation effects through the magnetosphere \citep[e.g.][]{bp12}. But invoking unusual physical processes may not be the most appealing option, as one might argue that the 12-second rotation period of PSR J2251$-$3711 is not exceptional enough to justify why such processes would have not been observed before in other pulsars. 

\section{Search for an X-ray Counterpart}\label{sec:xray}

\subsection{Archival data}

PSR~J2251$-$3711 has similar spin-down parameters to those of X-ray Isolated Neutron Stars (XINSs; see Figure~\ref{fig:ppdot}), which show soft thermal X-ray emission and have spin periods $P\sim$~3--17~s. We therefore searched archival catalogues for an X-ray counterpart. The only potential match that we found is 2RXS~J225144.6$-$371317 in the second \textit{ROSAT} all-sky survey source catalogue \citep{Boller+2016}, at a sky position of $\alpha$=22:51:44.69, $\delta$=$-$37:13:17.8. The reported detection likelihood is 10.05, corresponding to a non-negligible probability of spurious detection of 14\% \citep[Table 1 of][]{Boller+2016}, and the reported source count rate is $(5.8 \pm 2.0) \times 10^{-2}~\mathrm{s}^{-1}$. This candidate source lies 1\farcm5 away from the radio pulsar. The typical 1-$\sigma$ position uncertainty of sources in the \textit{ROSAT} catalogue is on the order of 0\farcm3, which suggests that 2RXS~J225144.6$-$371317 is unlikely to be related to PSR~J2251$-$3711.

For an absorbed blackbody model, the best-fit temperature reported in the catalogue for this candidate source is $kT = 22 \pm 3000$~eV; the disproportionately large uncertainty suggests that the fitting procedure failed and that the output value should not be trusted. The best-fit hydrogen column density $N_\mathrm{H} = (2.2 \pm 160) \times 10^{20} \,\mathrm{cm}^{-2}$ suffers the same problem, but at least appears consistent with what is expected from the DM$-N_\mathrm{H}$ relationship of \citet{hnk13} and a hypothetical source DM = $7 ~\mathrm{pc~cm}^{-3}$ (entirely reasonable for this line of sight). We can trust however that the spectral shape of the \textit{ROSAT} source is quite soft, since all 19 source counts were detected in the lowest energy band 100$-$440 eV. This is consistent with thermal emission from an XINS, of which all known specimens have blackbody temperatures $kT$ between 50 and 107 eV \citep[Table 3 of][]{Vigano+2013}. Converting\footnote{For such conversion purposes, we used the \textit{WebPIMMS} mission count rate simulator throughout this section: \url{https://heasarc.gsfc.nasa.gov/cgi-bin/Tools/w3pimms/w3pimms.pl}} the \textit{ROSAT} count rate into an unabsorbed bolometric luminosity yields $L_{\mathrm{2RXS}}^{85~\mathrm{eV}} = 1.2 \times 10^{32}\,(d / 1~\mathrm{kpc})^2~\mathrm{erg/s}$, where $d$ is the distance to the source and where we have assumed $kT = 85$~eV, the average XINS temperature. This is compatible with known XINS luminosities \citep{Vigano+2013}, providing additional incentive to re-observe the field.

\subsection{\textit{Swift} Observations}

\begin{figure*}
\includegraphics[width=0.75\textwidth]{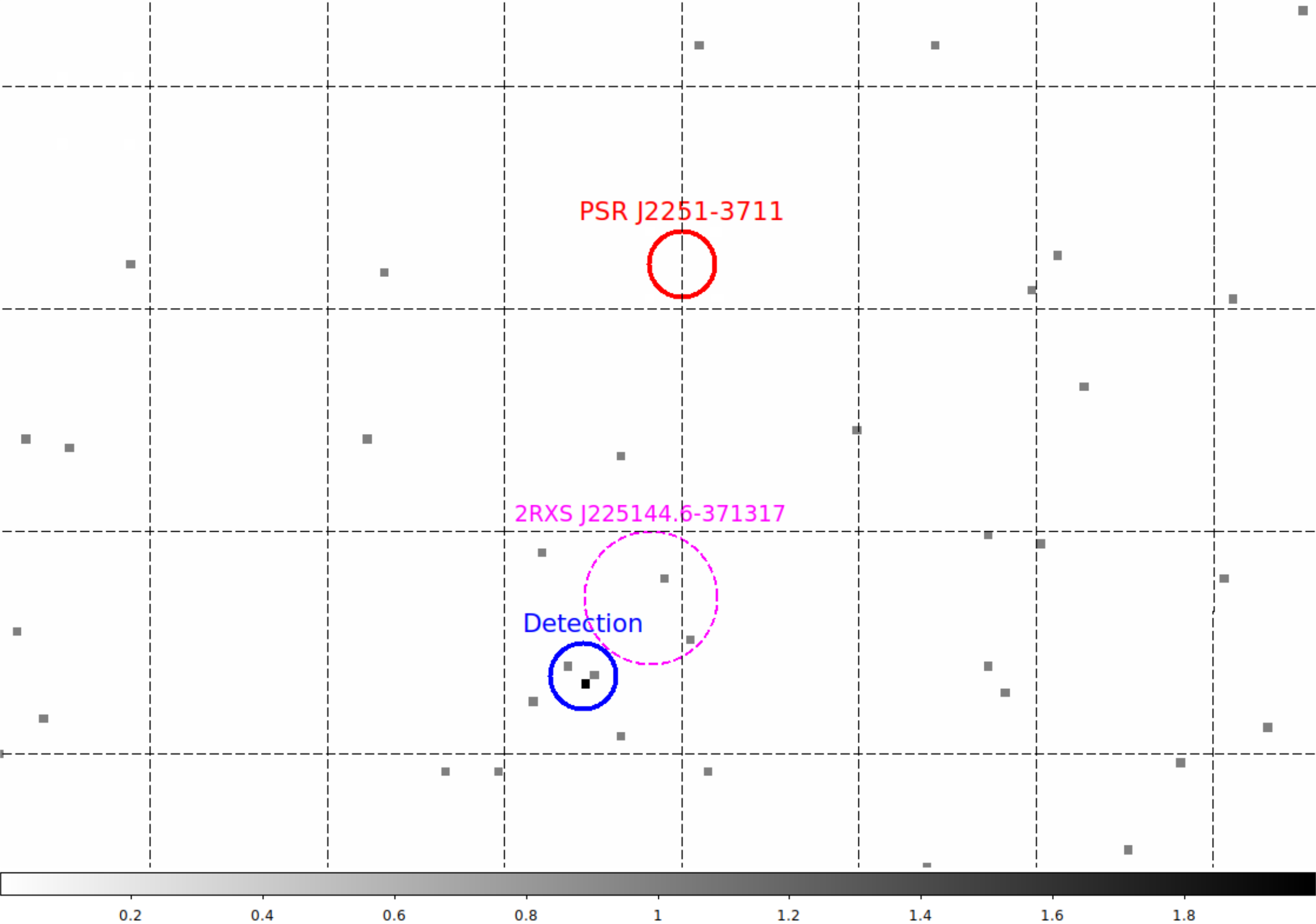}
\centering
\caption{Swift X-ray image of the field, with a total exposure time of 4.4 ks. Only a zoom on the region of interest is shown. The coordinate grid spacing is 1\arcmin~on both RA (horizontal) and Dec (vertical) axes. Red circle: half power diameter (18\arcsec) of the Swift PSF centered on the radio timing position of PSR~J2251$-$3711, enclosing zero counts. Blue circle: a 4-count detection with a statistical significance of 4.5-$\sigma$. Magenta dashed circle: the 1-$\sigma$ position uncertainty on the original detection of the \textit{ROSAT} faint source 2RXS~J225144.6$-$371317.}
\label{fig:swift_exposure}
\end{figure*}

In an attempt to detect the pulsar in the soft X-ray band, we obtained a total of $T_{\mathrm{obs}}=4.4$ ks of exposure time with the \textit{Neil Gehrels Swift Observatory} X-Ray Telescope (\textit{Swift} XRT), which covers the 0.3$-$10 keV band. The observations were taken on May 5th and 8th 2019 (Target ID: 00011329), with the field centered on the position of the \textit{ROSAT} source discussed above. We stacked the observations and analysed the resulting image, shown in Figure \ref{fig:swift_exposure}. Around the pulsar's timing position (Table \ref{tab:ephemeris}), no counts were detected within the 9\arcsec~ half-power radius of the \textit{Swift} XRT point-spread function (PSF). From this we can infer, using Table 3 of \citet{kbn91}, an upper bound at the 99\% confidence level on the total source counts (within the PSF) $\lambda_{S,u} = 4.6$, and an associated count rate $R_{S,u} = \lambda_{S,u} / T_{\mathrm{obs}} = 1.03 \times 10^{-3}~\mathrm{s}^{-1}$. In order to convert to an unabsorbed bolometric thermal luminosity, we need an estimate of the hydrogen column density, which we set to $N_\mathrm{H} = 3.7 \times 10^{20}\,\mathrm{cm}^{-2}$ based on the DM$-N_\mathrm{H}$ relationship of \citet{hnk13}. Assuming a blackbody source model with the average temperature of known XINS $kT = 85~\mathrm{eV}$, $R_{S,u}$ corresponds to an unabsorbed bolometric thermal luminosity of $L_{S,u}^{85 \mathrm{eV}} = 1.1 \times 10^{31}\,(d / 1~\mathrm{kpc})^2~\mathrm{erg/s}$ where $d$ is the distance to the pulsar. This luminosity (at $d = 1$ kpc) corresponds to the lower end of the XINS luminosity distribution. We note however that the upper bound thus derived is sensitive to the postulated source temperature; using the lowest reported XINS temperature ($kT = 50\,\mathrm{eV}$) instead yields a value 5 times larger $L_{S,u}^{50 \mathrm{eV}} = 5.3 \times 10^{31}\,(d / 1~\mathrm{kpc})^2~\mathrm{erg/s}$, close to the median XINS luminosity.

In this \textit{Swift} exposure, it is interesting to note that a source is detected with a significance level of $4.5\sigma$ (4 counts) within 35\arcsec\ of the position of 2RXS~J225144.6$-$371317. Given the \textit{ROSAT} positional uncertainty of 20\arcsec\ and the \textit{Swift} XRT PSF half-power radius of 9\arcsec, attributing both detections to the same underlying source is tempting. However, we first need to examine our second set of X-ray observations before attempting a fully informed interpretation of the field (see \S \ref{subsec:field_interpretation}).

\subsection{NICER Observations}

We also performed follow-up observations of PSR~J2251$-$3711 with the \textit{Neutron star Interior Composition Explorer} (\textit{NICER}), an X-ray observatory attached in 2017 to the International Space Station. The \textit{NICER} X-ray timing instrument is non-imaging, has a field of view of 6\arcmin\ and covers the soft X-ray energy range 0.2--12~keV with a large effective area (1900 cm$^2$ at 1.5~keV) and high time resolution ($\lesssim$100~ns). It is therefore well suited to the observation of potential XINS. Observations were taken from 2017 December 18 through 2018 June 14, spanning observation IDs 1020650101--102065014 and a net exposure time of 62~ks. For spectroscopic investigations using the non-imaging detectors, the \textit{NICER} team have developed two different background modeling methods: the space-weather (SW) model and 3C50 model. We used optimized filtering criteria for each model to select good time intervals of low particle and optical loading backgrounds and exclude noisy detector modules. After this filtering process, the X-ray spectra showed weak residuals below 1~keV using either background model. The filtered and background-subtracted 0.3--1~keV count rates were 0.019(3) and 0.056(4) counts~s$^{-1}$, corresponding to 7\% and 23\% of the background count rate for the SW and 3C50 background models, respectively. 

Fitting the time-averaged, background-subtracted \textit{NICER} spectra with an absorbed blackbody model, we found temperatures $kT=86\pm11$~eV and $96\pm 7$~eV, emission radii $R=0.84\mbox{ km }(d/1~{\rm kpc})$ and $R=1.0\mbox{ km }(d/1~{\rm kpc})$ for the SW and 3C50 background models, respectively. These correspond to absorbed X-ray fluxes in the 0.3--1 keV band of $2.0^{+0.1}_{-1.3}\times 10^{-14}$~erg~s$^{-1}$~cm$^{-2}$ and $5.6^{+0.8}_{-0.7}\times 10^{-14}$~erg~s$^{-1}$~cm$^{-2}$ respectively. Converting them to unabsorbed luminosities at $d = 1$~kpc assuming the hydrogen column density previously postulated for the pulsar yields $7.7^{+0.4}_{-5.0} \times 10^{30}$~erg~s$^{-1}$~cm$^{-2}$ and $1.8^{+0.3}_{-0.2} \times 10^{31}$~erg~s$^{-1}$~cm$^{-2}$ respectively. The first value is compatible with the 99\% confidence upper bound previously placed on PSR~J2251$-$3711's luminosity for a similar temperature of $kT = 85$~eV, but the second is about 60\% larger, suggesting that the majority of the \textit{NICER} counts do \textit{not} originate from the pulsar.

Nonetheless, we searched barycentric corrected X-ray events for pulsations at the spin period measured for PSR~J2251$-$3711 from radio timing data. We did not find any significant signal in the 0.3--1.0~keV band. After subtraction of background contributions assuming the SW and 3C50 models, 3$\sigma$ upper limits on the intrinsic source pulsed fraction were estimated to be 100\% and 88\% for these two models, respectively. In the former case, the low source count rate makes any pulsation undetectable even if the emission was 100\% pulsed.

\subsection{Interpreting the field}
\label{subsec:field_interpretation}

Overall there are three tentative X-ray band detections in the field: an integrated \textit{NICER} spectrum, a \textit{ROSAT} catalogue source and a 4-count cluster in a 4.4-ks \textit{Swift} exposure. All three are of limited statistical significance, which precludes any categorical interpretations of the data. Also, due to the relatively large 3\arcmin\ radius of the \textit{NICER} field of view, the provenance of the photons it collected needs to be considered. We can at least rule out the possibility of 2RXS~J225144.6$-$371317 being a steady thermal source with a temperature in the range of known XINS. Indeed, the luminosity inferred from the \textit{ROSAT} source count rate lies far above the detection thresholds of our observations; the source should have manifested itself in the \textit{Swift} observation as dozens of counts and registered a detectable blackbody spectrum in any individual \textit{NICER} exposure. It remains possible to reconcile the parameters reported by ROSAT with the absence of a clear \textit{Swift} detection, if we assume that its temperature kT is lower than 35 eV. Otherwise, 2RXS~J225144.6$-$371317 is either a variable source that remained in a quiet state during our observation campaign (December 2017 to May 2019), or the original \textit{ROSAT} detection was spurious.

The 4-count detection in the \textit{Swift} exposure could be a more interesting case, being coincident to 4\arcsec\ with a star of magnitude 21 in the Gaia data release 2 \citep[][Source ID: 6547998120327478272]{GAIADR2}. No further data (e.g., parallax or spectral type) are currently available for this object. If it is responsible for the X-ray counts, then it would likely be too bright in the optical to be an isolated neutron star, and too faint in the X-ray band to be part of an X-ray binary.

In any case, no candidate source in the field appears to be bright enough to single-handedly account for the \textit{NICER} blackbody-shaped spectrum. The most reasonable explanation is that the \textit{NICER} spectrum originates from other background sources. The \textit{Swift} observation currently provides the best constraint on PSR~J2251$-$3711's X-ray luminosity.

\section{Discussion}\label{sec:disc}

\subsection{Nature of PSR~J2251-3711}

Considering the position of PSR~J2251$-$3711 in the $P - \dot{P}$ diagram, its possible relationship to X-ray emitting neutron star classes deserves to be examined. We can certainly exclude that PSR~J2251$-$3711 is an active magnetar, as the typical X-ray luminosity of such an object \citep[$10^{33} - 10^{36}$ erg/s,][]{ok14} would have been easily detected in our data; but the idea of PSR~J2251$-$3711 being a low-B magnetar similar to SGR~0418$+$5729 \citep{ret+10} or Swift~J1822.3$-$1606 \citep{lsk+11} seems plausible at first sight. All three objects share very similar spin characteristics, which is evident from Figure \ref{fig:ppdot}. SGR~0418$+$5729 could have credibly remained undetected in our observations (even placed at a 1~kpc distance) given its X-ray flux in quiescence of less than a few $10^{-14}~\mathrm{erg~cm}^{-2}~\mathrm{s}^{-1}$ \citep{rip+13}, which would be consistent with the non-detection of our pulsar. However, most magnetars are either radio quiet, or show particularly wide radio profiles with duty cycles in excess of 10\% with significant time-variability, which may include weeks to months of nulling \citep{cjh+08, lbb+10}. This makes a magnetar classification for PSR~J2251$-$3711 much less credible. Furthermore, its high Galactic latitude ($b = -62.9 \degree$) stands in stark contrast to the magnetar population, although we will expand on this specific point below.

On the other hand, there are no strong arguments against PSR~J2251$-$3711 being an XINS. Its period, period derivative, Galactic latitude and distance all appear compatible with that of known thermally-emitting isolated neutron stars. The main question here is of course the non-detection of the pulsar in the X-ray band. The 99\% confidence upper bound on its bolometric unabsorbed thermal luminosity, derived in the previous section, is $L_{S,u}^{85 \mathrm{eV}} = 1.1 \times 10^{31} ~\mathrm{erg/s}$ assuming that it lies at a distance of 1 kpc. This luminosity is comparable to that of RX~J0420.0$-$5022 and RX~J1605.3$+$3249, the two faintest known XINS. One must however take into account a margin of error of nearly an order of magnitude on this upper bound, due to uncertainties on the distance to the PSR~J2251$-$3711 and its unknown temperature. Furthermore, we can consider its X-ray over spin-down luminosity ratio $L_X / \dot{E}$, which provides another useful means of distinguishing between neutron star classes; see e.g., Figure 12 of \citet{Enoto2019} for an up-to-date $L_X - \dot{E}$ diagram of the neutron star population. For all reasonable distance and temperature estimates, the upper bound on $L_X / \dot{E}$ for PSR~J2251$-$3711 remains larger than 1, compared to a value of $\lesssim 0.01$ that would be required to confidently place it in into the rotation-powered pulsar category. It is therefore currently not possible to rule out an XINS nature for PSR~J2251$-$3711. The matter will only be settled with observations using sensitive X-ray observatories with imaging capabilities, such as \textit{XMM-Newton} or the \textit{Chandra X-ray Observatory}.

\subsection{Evolution history of PSR~J2251-3711}

There are two plausible and quite distinct evolution scenarios for PSR~J2251$-$3711 that we shall discuss here. The first is that it followed the standard picture, where it would have been born with parameters similar to those of the Crab pulsar and have undergone a spin-down evolution dominated by magnetic dipole braking. In this case, its true age should be comparable to its characteristic age of ~15 Myr. There are however significant caveats to the dipole braking model, especially when the task at hand is to determine the age of a radio pulsar. Firstly, the model assumes that the inclination angle $\alpha$ between the spin and magnetic axes remains constant, while there is some observational evidence that both axes tend to align over time \citep{tm98}. Such decrease of $\alpha$ manifests itself as an increased braking index $n$ \citep{tk01}, potentially giving the illusion of a decreasing characteristic surface magnetic field if $n > 3$. Secondly, the accurately known braking indices (noted $n$ hereafter) for young pulsars are in the range $-$1.2 to 3.2 \citep{els17, agf+16}; this suggests that pulsars are born with $n < 3$, which corresponds to an increase of the characteristic surface magnetic field \citep[either apparent or possibly physical,][]{elk+11, Ho2015}. Taking into account these two extra ingredients, namely a plausible distribution of braking indices at birth and a negative $\dot{\alpha}$ term, \citet{JK2017} managed to reproduce the bulk of the $P$--$\dot{P}$ diagram population of isolated NS. Interestingly, their model postulated that the intrinsic magnetic field of pulsars does \textit{not} decay. Among their conclusions was that the characteristic age is then a systematic overestimate of the true age. We can infer from their Figure 5 that a pulsar with a characteristic age of 15 Myr born in the same $P$--$\dot{P}$ region as the Crab should see its true age estimate reduced by half an order of magnitude, to approximately 4 Myr. 

The alternative evolution scenario for PSR~J2251$-$3711 is that it started its neutron star life as a magnetar. It is generally accepted that the high energy emission of a magnetar is powered by the dissipation of its magnetic field; in strongly magnetized neutron stars ($B \gtrsim 10^{13}$~G), it has been shown that the time evolution of the temperature and magnetic field of the star strongly depend on each other and must be treated simultaneously \citep{apm08, pmg09}. This has sparked the development of so-called magneto-thermal evolution models of neutron stars, the most advanced being that of \citet{Vigano+2013} which provides a number of directly testable predictions, in particular the trajectory that a neutron star follows in the $P$--$\dot{P}$ diagram as it cools down. The amount of magnetic field decay over time depends strongly on the initial magnetic field configuration postulated for the star, which could be exclusively crustal (their model A), or with a significant core component (their models B and C). In the latter case, the dipole component of the magnetic field is shown to remain approximately constant, and there $n = 3$ braking is expected as above, also corresponding to an age of several Myr. But an initially crustal field decays significantly over time, which is accompanied by a rapid spin-down as shown on their Figure 10. It therefore appears plausible that PSR~J2251$-$3711 was born as ``model A" magnetar with an initial dipolar surface magnetic field strength $B \simeq 3 \times 10^{14}$G, which, if we are to trust the model of \citet{Vigano+2013}, would make it approximately 0.4 Myr old.

Choosing between evolution models with or without magnetic field decay for PSR~J2251$-$3711 clearly rests upon obtaining an estimate of its true age, given that they predict ages that differ by an order of magnitude. The detection of an X-ray counterpart would be direct evidence that the pulsar is younger than the typical NS cooling time of 1 Myr and would strongly argue in favour of a magnetar origin. The unusually high Galactic latitude of PSR~J2251$-$3711 might also provide another means of estimating its true age; if we assume that its parent supernova occurred in the Galactic plane, then a lower bound on the vertical component of its kick velocity would be

\begin{equation}
v_z = 870 \left( \frac{d}{\mathrm{kpc}} \right) \left( \frac{T}{\mathrm{Myr}} \right)^{-1} \mathrm{km/s},
\end{equation}
where $d$ is the present distance to the pulsar and $T$ its age. The velocity component transverse to our line of sight would have approximately half ($\cos{b} = 0.46$) the value above. The resulting observable proper motion further away from the Galactic plane would then be

\begin{equation}
|\dot{b}| = 84 \left( \frac{T}{\mathrm{Myr}} \right)^{-1} \mathrm{mas/yr}.
\end{equation}
The young age of $\approx 0.4$ Myr predicted by model A of \citet{Vigano+2013} may therefore manifest itself through a proper motion close to the highest values currently known for a radio pulsar \citep{PSRCAT}. Timing at 1.4~GHz is unlikely to ever yield a measurable proper motion due to the high timing residuals of the source, which currently limits positional accuracy to a few arcseconds; long baseline radio interferometry is therefore required. An alternative would be to estimate the velocity of PSR~J2251$-$3711 relative to the interstellar medium from measurements of its diffractive scintillation time and frequency scales \citep[e.g.,][]{cr98, jnk98}; the source is not bright enough in our radio data to make such a measurement practical, but the prospects of doing so with a more sensitive facility such as MeerKAT are very good. We note as a caveat to this discussion that a nearby neutron star such as PSR~J2251$-$3711 may have been born significantly out of the Galactic plane, due to the presence of a significant number of potential progenitors off the plane within a 1 kpc radius -- the so-called OB runaway stars \citep{pph+08}.

In summary, it is possible to find evidence that PSR~J2251$-$3711 is young, but no such direct avenues exist to demonstrate that it is old. Confirmation of old age would instead likely to be found in a persistent lack of evidence for youth in future observations, in both radio and X-ray bands.

\section{Conclusion}\label{sec:conc}

We have presented radio and X-ray observations of PSR~J2251$-$3711, a newly discovered radio pulsar with an unusually long spin period of 12.1 seconds, the second largest known. Its radio emission is intermittent with a $\sim$65\% nulling fraction, but it does not appear to shut down on timescales of hours or days. It shows a small sample of single pulses with 180-degree sweeps of the polarization position angle, that remarkably occurs at different phases from one pulse to the other. This observation cannot be easily reconciled with the rotating vector model, but we have suggested a few tentative explanations to be further explored. We have also shown that PSR~J2251$-$3711 is unlikely to be a low-B magnetar; however, the possibility of it being a cooling X-ray isolated neutron star (XINS) remains open, which must be tested with deeper X-ray imaging observations.

PSR~J2251$-$3711 lies in a region of the $P - \dot{P}$ diagram predicted to be a magnetar graveyard by magneto-thermal evolution models, but that same region could also credibly contain ordinary (and much older) radio pulsars that underwent a spin-down evolution with little to no magnetic field decay. It will be interesting in the near future to determine which evolutionary paths the emerging population of very slow radio pulsars ($P > 10$~s) actually followed. This should hopefully bring new constraints to magneto-thermal evolution models and overall contribute to a unified vision of the apparent neutron star diversity.

\section*{Acknowledgements}

VM, BWS, MC and FJ acknowledge funding from the European Research Council (ERC) under the European Union's Horizon 2020 research and innovation programme (grant agreement No. 694745). SG acknowledges the support of the CNES. The Parkes Observatory is part of the Australia Telescope which is funded by the Commonwealth of Australia for operation as a National Facility managed by CSIRO. This research is supported by the Australian Research Council through grants FL150100148 and CE170100004. Some of this work was performed on the OzSTAR national facility at Swinburne University of Technology. OzSTAR is funded by Swinburne and the National Collaborative Research Infrastructure Strategy (NCRIS). The NICER mission and portions of the NICER science team activities are funded by NASA. VM thanks Kostas Gourgouliatos, Sergei Popov and Aris Karastergiou for useful discussions, and the organisers of the PHAROS conference ``the multi-messenger physics and astrophysics of neutron stars" for enabling these discussions in the first place. We also thank Christina Ilie and Patrick Weltevrede for their valuable guidance on single pulse analysis and useful discussions on the single pulse polarization behaviour of PSR J2251$-$3711. Finally, we thank the anonymous referee for providing valuable comments.



\bibliographystyle{mnras}
\bibliography{references}

\begin{thebibliography}{}
\makeatletter
\relax
\def\mn@urlcharsother{\let\do\@makeother \do\$\do\&\do\#\do\^\do\_\do\%\do\~}
\def\mn@doi{\begingroup\mn@urlcharsother \@ifnextchar [ {\mn@doi@}
  {\mn@doi@[]}}
\def\mn@doi@[#1]#2{\def\@tempa{#1}\ifx\@tempa\@empty \href
  {http://dx.doi.org/#2} {doi:#2}\else \href {http://dx.doi.org/#2} {#1}\fi
  \endgroup}
\def\mn@eprint#1#2{\mn@eprint@#1:#2::\@nil}
\def\mn@eprint@arXiv#1{\href {http://arxiv.org/abs/#1} {{\tt arXiv:#1}}}
\def\mn@eprint@dblp#1{\href {http://dblp.uni-trier.de/rec/bibtex/#1.xml}
  {dblp:#1}}
\def\mn@eprint@#1:#2:#3:#4\@nil{\def\@tempa {#1}\def\@tempb {#2}\def\@tempc
  {#3}\ifx \@tempc \@empty \let \@tempc \@tempb \let \@tempb \@tempa \fi \ifx
  \@tempb \@empty \def\@tempb {arXiv}\fi \@ifundefined
  {mn@eprint@\@tempb}{\@tempb:\@tempc}{\expandafter \expandafter \csname
  mn@eprint@\@tempb\endcsname \expandafter{\@tempc}}}

\bibitem[\protect\citeauthoryear{{Aguilera}, {Pons}  \& {Miralles}}{{Aguilera}
  et~al.}{2008}]{apm08}
{Aguilera} D.~N.,  {Pons} J.~A.,   {Miralles} J.~A.,  2008, \mn@doi [\apjl]
  {10.1086/527547}, \href
  {https://ui.adsabs.harvard.edu/abs/2008ApJ...673L.167A} {673, L167}

\bibitem[\protect\citeauthoryear{{Archibald} et~al.,}{{Archibald}
  et~al.}{2016}]{agf+16}
{Archibald} R.~F.,  et~al., 2016, \mn@doi [\apjl]
  {10.3847/2041-8205/819/1/L16}, \href
  {https://ui.adsabs.harvard.edu/abs/2016ApJ...819L..16A} {819, L16}

\bibitem[\protect\citeauthoryear{{Beskin} \& {Philippov}}{{Beskin} \&
  {Philippov}}{2012}]{bp12}
{Beskin} V.~S.,  {Philippov} A.~A.,  2012, \mn@doi [\mnras]
  {10.1111/j.1365-2966.2012.20988.x}, \href
  {https://ui.adsabs.harvard.edu/abs/2012MNRAS.425..814B} {425, 814}

\bibitem[\protect\citeauthoryear{{Bhattacharya} \& {van den
  Heuvel}}{{Bhattacharya} \& {van den Heuvel}}{1991}]{bh91}
{Bhattacharya} D.,  {van den Heuvel} E.~P.~J.,  1991, \mn@doi [\physrep]
  {10.1016/0370-1573(91)90064-S}, \href
  {https://ui.adsabs.harvard.edu/abs/1991PhR...203....1B} {203, 1}

\bibitem[\protect\citeauthoryear{{Blaskiewicz}, {Cordes}  \&
  {Wasserman}}{{Blaskiewicz} et~al.}{1991}]{bcw91}
{Blaskiewicz} M.,  {Cordes} J.~M.,   {Wasserman} I.,  1991, \mn@doi [\apj]
  {10.1086/169850}, \href
  {https://ui.adsabs.harvard.edu/abs/1991ApJ...370..643B} {370, 643}

\bibitem[\protect\citeauthoryear{{Boller}, {Freyberg}, {Tr{\"u}mper}, {Haberl},
  {Voges}  \& {Nandra}}{{Boller} et~al.}{2016}]{Boller+2016}
{Boller} T.,  {Freyberg} M.~J.,  {Tr{\"u}mper} J.,  {Haberl} F.,  {Voges} W.,
  {Nandra} K.,  2016, \mn@doi [\aap] {10.1051/0004-6361/201525648}, \href
  {http://adsabs.harvard.edu/abs/2016A%26A...588A.103B} {588, A103}

\bibitem[\protect\citeauthoryear{{Burrows} et~al.,}{{Burrows}
  et~al.}{2005}]{Burrows2005}
{Burrows} D.~N.,  et~al., 2005, \mn@doi [\ssr] {10.1007/s11214-005-5097-2},
  \href {https://ui.adsabs.harvard.edu/abs/2005SSRv..120..165B} {120, 165}

\bibitem[\protect\citeauthoryear{{Camilo}, {Reynolds}, {Johnston}, {Halpern}
  \& {Ransom}}{{Camilo} et~al.}{2008}]{cjh+08}
{Camilo} F.,  {Reynolds} J.,  {Johnston} S.,  {Halpern} J.~P.,   {Ransom}
  S.~M.,  2008, \mn@doi [\apj] {10.1086/587054}, \href
  {https://ui.adsabs.harvard.edu/abs/2008ApJ...679..681C} {679, 681}

\bibitem[\protect\citeauthoryear{{Camilo}, {Ransom}, {Chatterjee}, {Johnston}
  \& {Demorest}}{{Camilo} et~al.}{2012}]{crc+12}
{Camilo} F.,  {Ransom} S.~M.,  {Chatterjee} S.,  {Johnston} S.,   {Demorest}
  P.,  2012, \mn@doi [ApJ] {10.1088/0004-637X/746/1/63}, \href
  {https://ui.adsabs.harvard.edu/#abs/2012ApJ...746...63C} {746, 63}

\bibitem[\protect\citeauthoryear{{Chen} \& {Ruderman}}{{Chen} \&
  {Ruderman}}{1993}]{cr93}
{Chen} K.,  {Ruderman} M.,  1993, \mn@doi [\apj] {10.1086/172129}, \href
  {https://ui.adsabs.harvard.edu/abs/1993ApJ...402..264C} {402, 264}

\bibitem[\protect\citeauthoryear{{Cordes} \& {Lazio}}{{Cordes} \&
  {Lazio}}{2002}]{NE2001}
{Cordes} J.~M.,  {Lazio} T.~J.~W.,  2002, arXiv e-prints, \href
  {https://ui.adsabs.harvard.edu/abs/2002astro.ph..7156C} {pp
  astro--ph/0207156}

\bibitem[\protect\citeauthoryear{{Cordes} \& {Rickett}}{{Cordes} \&
  {Rickett}}{1998}]{cr98}
{Cordes} J.~M.,  {Rickett} B.~J.,  1998, \mn@doi [\apj] {10.1086/306358}, \href
  {https://ui.adsabs.harvard.edu/abs/1998ApJ...507..846C} {507, 846}

\bibitem[\protect\citeauthoryear{{Deller} et~al.,}{{Deller}
  et~al.}{2019}]{dgb+19}
{Deller} A.~T.,  et~al., 2019, \mn@doi [\apj] {10.3847/1538-4357/ab11c7}, \href
  {https://ui.adsabs.harvard.edu/abs/2019ApJ...875..100D} {875, 100}

\bibitem[\protect\citeauthoryear{Enoto, Kisaka  \& Shibata}{Enoto
  et~al.}{2019}]{Enoto2019}
Enoto T.,  Kisaka S.,   Shibata S.,  2019, \mn@doi [Reports on Progress in
  Physics] {10.1088/1361-6633/ab3def}, 82, 106901

\bibitem[\protect\citeauthoryear{{Espinoza}, {Lyne}, {Kramer}, {Manchester}  \&
  {Kaspi}}{{Espinoza} et~al.}{2011}]{elk+11}
{Espinoza} C.~M.,  {Lyne} A.~G.,  {Kramer} M.,  {Manchester} R.~N.,   {Kaspi}
  V.~M.,  2011, \mn@doi [\apjl] {10.1088/2041-8205/741/1/L13}, \href
  {https://ui.adsabs.harvard.edu/abs/2011ApJ...741L..13E} {741, L13}

\bibitem[\protect\citeauthoryear{{Espinoza}, {Lyne}  \& {Stappers}}{{Espinoza}
  et~al.}{2017}]{els17}
{Espinoza} C.~M.,  {Lyne} A.~G.,   {Stappers} B.~W.,  2017, \mn@doi [\mnras]
  {10.1093/mnras/stw3081}, \href
  {https://ui.adsabs.harvard.edu/abs/2017MNRAS.466..147E} {466, 147}

\bibitem[\protect\citeauthoryear{{Foreman-Mackey}, {Hogg}, {Lang}  \&
  {Goodman}}{{Foreman-Mackey} et~al.}{2013}]{emcee}
{Foreman-Mackey} D.,  {Hogg} D.~W.,  {Lang} D.,   {Goodman} J.,  2013, \mn@doi
  [Publications of the Astronomical Society of the Pacific] {10.1086/670067},
  \href {https://ui.adsabs.harvard.edu/\#abs/2013PASP..125..306F} {125, 306}

\bibitem[\protect\citeauthoryear{{Gaensler} \& {Frail}}{{Gaensler} \&
  {Frail}}{2000}]{gf00}
{Gaensler} B.~M.,  {Frail} D.~A.,  2000, \nat, \href
  {https://ui.adsabs.harvard.edu/abs/2000Natur.406..158G} {406, 158}

\bibitem[\protect\citeauthoryear{{Gaia Collaboration} et~al.,}{{Gaia
  Collaboration} et~al.}{2018}]{GAIADR2}
{Gaia Collaboration} et~al., 2018, \mn@doi [\aap]
  {10.1051/0004-6361/201833051}, \href
  {https://ui.adsabs.harvard.edu/abs/2018A&A...616A...1G} {616, A1}

\bibitem[\protect\citeauthoryear{{He}, {Ng}  \& {Kaspi}}{{He}
  et~al.}{2013}]{hnk13}
{He} C.,  {Ng} C.~Y.,   {Kaspi} V.~M.,  2013, \mn@doi [\apj]
  {10.1088/0004-637X/768/1/64}, \href
  {https://ui.adsabs.harvard.edu/abs/2013ApJ...768...64H} {768, 64}

\bibitem[\protect\citeauthoryear{{Hessels}, {Ransom}, {Stairs}, {Freire},
  {Kaspi}  \& {Camilo}}{{Hessels} et~al.}{2006}]{hrs+06}
{Hessels} J. W.~T.,  {Ransom} S.~M.,  {Stairs} I.~H.,  {Freire} P. C.~C.,
  {Kaspi} V.~M.,   {Camilo} F.,  2006, \mn@doi [Science]
  {10.1126/science.1123430}, \href
  {https://ui.adsabs.harvard.edu/#abs/2006Sci...311.1901H} {311, 1901}

\bibitem[\protect\citeauthoryear{{Ho}}{{Ho}}{2015}]{Ho2015}
{Ho} W. C.~G.,  2015, \mn@doi [\mnras] {10.1093/mnras/stv1339}, \href
  {https://ui.adsabs.harvard.edu/abs/2015MNRAS.452..845H} {452, 845}

\bibitem[\protect\citeauthoryear{{Hobbs}, {Edwards}  \& {Manchester}}{{Hobbs}
  et~al.}{2006}]{TEMPO2}
{Hobbs} G.~B.,  {Edwards} R.~T.,   {Manchester} R.~N.,  2006, \mn@doi [\mnras]
  {10.1111/j.1365-2966.2006.10302.x}, \href
  {https://ui.adsabs.harvard.edu/abs/2006MNRAS.369..655H} {369, 655}

\bibitem[\protect\citeauthoryear{{Hotan}, {van Straten}  \&
  {Manchester}}{{Hotan} et~al.}{2004}]{PSRCHIVE}
{Hotan} A.~W.,  {van Straten} W.,   {Manchester} R.~N.,  2004, \mn@doi
  [Publications of the Astronomical Society of Australia] {10.1071/AS04022},
  \href {https://ui.adsabs.harvard.edu/\#abs/2004PASA...21..302H} {21, 302}

\bibitem[\protect\citeauthoryear{{Johnston} \& {Karastergiou}}{{Johnston} \&
  {Karastergiou}}{2017}]{JK2017}
{Johnston} S.,  {Karastergiou} A.,  2017, \mn@doi [\mnras]
  {10.1093/mnras/stx377}, \href
  {https://ui.adsabs.harvard.edu/abs/2017MNRAS.467.3493J} {467, 3493}

\bibitem[\protect\citeauthoryear{{Johnston}, {Nicastro}  \&
  {Koribalski}}{{Johnston} et~al.}{1998}]{jnk98}
{Johnston} S.,  {Nicastro} L.,   {Koribalski} B.,  1998, \mn@doi [\mnras]
  {10.1046/j.1365-8711.1998.01461.x}, \href
  {https://ui.adsabs.harvard.edu/abs/1998MNRAS.297..108J} {297, 108}

\bibitem[\protect\citeauthoryear{{Kaspi}}{{Kaspi}}{2010}]{Kaspi2010}
{Kaspi} V.~M.,  2010, \mn@doi [Proceedings of the National Academy of Science]
  {10.1073/pnas.1000812107}, \href
  {https://ui.adsabs.harvard.edu/abs/2010PNAS..107.7147K} {107, 7147}

\bibitem[\protect\citeauthoryear{{Kaspi} \& {Beloborodov}}{{Kaspi} \&
  {Beloborodov}}{2017}]{kb17}
{Kaspi} V.~M.,  {Beloborodov} A.~M.,  2017, \mn@doi [\araa]
  {10.1146/annurev-astro-081915-023329}, \href
  {https://ui.adsabs.harvard.edu/abs/2017ARA&A..55..261K} {55, 261}

\bibitem[\protect\citeauthoryear{{Kaspi} \& {Kramer}}{{Kaspi} \&
  {Kramer}}{2016}]{KaspiKramer2016}
{Kaspi} V.~M.,  {Kramer} M.,  2016, arXiv e-prints, \href
  {https://ui.adsabs.harvard.edu/abs/2016arXiv160207738K} {p. arXiv:1602.07738}

\bibitem[\protect\citeauthoryear{{Keane}}{{Keane}}{2018}]{SUPERBI}
{Keane} E.~F.~e.,  2018, \mn@doi [MNRAS] {10.1093/mnras/stx2126}, \href
  {http://adsabs.harvard.edu/abs/2018MNRAS.473..116K} {473, 116}

\bibitem[\protect\citeauthoryear{{Keane} \& {Kramer}}{{Keane} \&
  {Kramer}}{2008}]{KeaneKramer2008}
{Keane} E.~F.,  {Kramer} M.,  2008, \mn@doi [\mnras]
  {10.1111/j.1365-2966.2008.14045.x}, \href
  {https://ui.adsabs.harvard.edu/abs/2008MNRAS.391.2009K} {391, 2009}

\bibitem[\protect\citeauthoryear{{Keane}, {Ludovici}, {Eatough}, {Kramer},
  {Lyne}, {McLaughlin}  \& {Stappers}}{{Keane} et~al.}{2010}]{kle+10}
{Keane} E.~F.,  {Ludovici} D.~A.,  {Eatough} R.~P.,  {Kramer} M.,  {Lyne}
  A.~G.,  {McLaughlin} M.~A.,   {Stappers} B.~W.,  2010, \mn@doi [\mnras]
  {10.1111/j.1365-2966.2009.15693.x}, \href
  {https://ui.adsabs.harvard.edu/abs/2010MNRAS.401.1057K} {401, 1057}

\bibitem[\protect\citeauthoryear{{Keane}, {McLaughlin}, {Kramer}, {Stappers},
  {Bassa}, {Purver}  \& {Weltevrede}}{{Keane} et~al.}{2013}]{kmk+13}
{Keane} E.~F.,  {McLaughlin} M.~A.,  {Kramer} M.,  {Stappers} B.~W.,  {Bassa}
  C.~G.,  {Purver} M.~B.,   {Weltevrede} P.,  2013, \mn@doi [ApJ]
  {10.1088/0004-637X/764/2/180}, \href
  {https://ui.adsabs.harvard.edu/#abs/2013ApJ...764..180K} {764, 180}

\bibitem[\protect\citeauthoryear{{Keith} et~al.,}{{Keith}
  et~al.}{2010}]{kjs+10}
{Keith} M.~J.,  et~al., 2010, \mn@doi [\mnras]
  {10.1111/j.1365-2966.2010.17325.x}, \href
  {https://ui.adsabs.harvard.edu/abs/2010MNRAS.409..619K} {409, 619}

\bibitem[\protect\citeauthoryear{{Kondratiev}, {McLaughlin}, {Lorimer},
  {Burgay}, {Possenti}, {Turolla}, {Popov}  \& {Zane}}{{Kondratiev}
  et~al.}{2009}]{Kondratiev2009}
{Kondratiev} V.~I.,  {McLaughlin} M.~A.,  {Lorimer} D.~R.,  {Burgay} M.,
  {Possenti} A.,  {Turolla} R.,  {Popov} S.~B.,   {Zane} S.,  2009, \mn@doi
  [\apj] {10.1088/0004-637X/702/1/692}, \href
  {https://ui.adsabs.harvard.edu/abs/2009ApJ...702..692K} {702, 692}

\bibitem[\protect\citeauthoryear{{Kraft}, {Burrows}  \& {Nousek}}{{Kraft}
  et~al.}{1991}]{kbn91}
{Kraft} R.~P.,  {Burrows} D.~N.,   {Nousek} J.~A.,  1991, \mn@doi [\apj]
  {10.1086/170124}, \href
  {https://ui.adsabs.harvard.edu/abs/1991ApJ...374..344K} {374, 344}

\bibitem[\protect\citeauthoryear{{Kramer}, {Lyne}, {Hobbs}, {L{\"o}hmer},
  {Carr}, {Jordan}  \& {Wolszczan}}{{Kramer} et~al.}{2003}]{klh+03}
{Kramer} M.,  {Lyne} A.~G.,  {Hobbs} G.,  {L{\"o}hmer} O.,  {Carr} P.,
  {Jordan} C.,   {Wolszczan} A.,  2003, \mn@doi [\apjl] {10.1086/378082}, \href
  {https://ui.adsabs.harvard.edu/abs/2003ApJ...593L..31K} {593, L31}

\bibitem[\protect\citeauthoryear{{Kramer}, {Lyne}, {O'Brien}, {Jordan}  \&
  {Lorimer}}{{Kramer} et~al.}{2006}]{mlo+06}
{Kramer} M.,  {Lyne} A.~G.,  {O'Brien} J.~T.,  {Jordan} C.~A.,   {Lorimer}
  D.~R.,  2006, \mn@doi [Science] {10.1126/science.1124060}, \href
  {https://ui.adsabs.harvard.edu/#abs/2006Sci...312..549K} {312, 549}

\bibitem[\protect\citeauthoryear{{Levin} et~al.,}{{Levin}
  et~al.}{2010}]{lbb+10}
{Levin} L.,  et~al., 2010, \mn@doi [\apjl] {10.1088/2041-8205/721/1/L33}, \href
  {https://ui.adsabs.harvard.edu/abs/2010ApJ...721L..33L} {721, L33}

\bibitem[\protect\citeauthoryear{{Livingstone}, {Scholz}, {Kaspi}, {Ng}  \&
  {Gavriil}}{{Livingstone} et~al.}{2011}]{lsk+11}
{Livingstone} M.~A.,  {Scholz} P.,  {Kaspi} V.~M.,  {Ng} C.~Y.,   {Gavriil}
  F.~P.,  2011, \mn@doi [\apjl] {10.1088/2041-8205/743/2/L38}, \href
  {https://ui.adsabs.harvard.edu/abs/2011ApJ...743L..38L} {743, L38}

\bibitem[\protect\citeauthoryear{{Lyne} \& {Graham-Smith}}{{Lyne} \&
  {Graham-Smith}}{2012}]{lgs12}
{Lyne} A.,  {Graham-Smith} F.,  2012, {Pulsar Astronomy}

\bibitem[\protect\citeauthoryear{{Lyne}, {Jordan}, {Graham-Smith}, {Espinoza},
  {Stappers}  \& {Weltevrede}}{{Lyne} et~al.}{2015}]{Lyne2015}
{Lyne} A.~G.,  {Jordan} C.~A.,  {Graham-Smith} F.,  {Espinoza} C.~M.,
  {Stappers} B.~W.,   {Weltevrede} P.,  2015, \mn@doi [MNRAS]
  {10.1093/mnras/stu2118}, \href
  {http://adsabs.harvard.edu/abs/2015MNRAS.446..857L} {446, 857}

\bibitem[\protect\citeauthoryear{{Manchester}, {Taylor}  \&
  {Huguenin}}{{Manchester} et~al.}{1975}]{mth75}
{Manchester} R.~N.,  {Taylor} J.~H.,   {Huguenin} G.~R.,  1975, \mn@doi [\apj]
  {10.1086/153395}, \href
  {https://ui.adsabs.harvard.edu/abs/1975ApJ...196...83M} {196, 83}

\bibitem[\protect\citeauthoryear{{Manchester}, {Hobbs}, {Teoh}  \&
  {Hobbs}}{{Manchester} et~al.}{2005}]{PSRCAT}
{Manchester} R.~N.,  {Hobbs} G.~B.,  {Teoh} A.,   {Hobbs} M.,  2005, \mn@doi
  [\aj] {10.1086/428488}, \href
  {https://ui.adsabs.harvard.edu/abs/2005AJ....129.1993M} {129, 1993}

\bibitem[\protect\citeauthoryear{{Noutsos}, {Schnitzeler}, {Keane}, {Kramer}
  \& {Johnston}}{{Noutsos} et~al.}{2013}]{Noutsos2013}
{Noutsos} A.,  {Schnitzeler} D.~H.~F.~M.,  {Keane} E.~F.,  {Kramer} M.,
  {Johnston} S.,  2013, \mn@doi [MNRAS] {10.1093/mnras/stt047}, \href
  {https://ui.adsabs.harvard.edu/#abs/2013MNRAS.430.2281N} {430, 2281}

\bibitem[\protect\citeauthoryear{{Olausen} \& {Kaspi}}{{Olausen} \&
  {Kaspi}}{2014}]{ok14}
{Olausen} S.~A.,  {Kaspi} V.~M.,  2014, \mn@doi [\apjs]
  {10.1088/0067-0049/212/1/6}, \href
  {https://ui.adsabs.harvard.edu/abs/2014ApJS..212....6O} {212, 6}

\bibitem[\protect\citeauthoryear{{Patterson}}{{Patterson}}{1979}]{pat79}
{Patterson} J.,  1979, \mn@doi [\apj] {10.1086/157582}, \href
  {https://ui.adsabs.harvard.edu/abs/1979ApJ...234..978P} {234, 978}

\bibitem[\protect\citeauthoryear{{Pons}, {Miralles}  \& {Geppert}}{{Pons}
  et~al.}{2009}]{pmg09}
{Pons} J.~A.,  {Miralles} J.~A.,   {Geppert} U.,  2009, \mn@doi [\aap]
  {10.1051/0004-6361:200811229}, \href
  {https://ui.adsabs.harvard.edu/abs/2009A&A...496..207P} {496, 207}

\bibitem[\protect\citeauthoryear{{Posselt}, {Popov}, {Haberl}, {Tr{\"u}mper},
  {Turolla}  \& {Neuh{\"a}user}}{{Posselt} et~al.}{2008}]{pph+08}
{Posselt} B.,  {Popov} S.~B.,  {Haberl} F.,  {Tr{\"u}mper} J.,  {Turolla} R.,
  {Neuh{\"a}user} R.,  2008, \mn@doi [\aap] {10.1051/0004-6361:20078430}, \href
  {https://ui.adsabs.harvard.edu/abs/2008A&A...482..617P} {482, 617}

\bibitem[\protect\citeauthoryear{{Potekhin}, {Pons}  \& {Page}}{{Potekhin}
  et~al.}{2015}]{ppp15}
{Potekhin} A.~Y.,  {Pons} J.~A.,   {Page} D.,  2015, \mn@doi [\ssr]
  {10.1007/s11214-015-0180-9}, \href
  {https://ui.adsabs.harvard.edu/abs/2015SSRv..191..239P} {191, 239}

\bibitem[\protect\citeauthoryear{{Radhakrishnan} \& {Cooke}}{{Radhakrishnan} \&
  {Cooke}}{1969}]{rc69}
{Radhakrishnan} V.,  {Cooke} D.~J.,  1969, \aplett, \href
  {https://ui.adsabs.harvard.edu/abs/1969ApL.....3..225R} {3, 225}

\bibitem[\protect\citeauthoryear{{Rea} et~al.,}{{Rea} et~al.}{2010}]{ret+10}
{Rea} N.,  et~al., 2010, \mn@doi [Science] {10.1126/science.1196088}, \href
  {https://ui.adsabs.harvard.edu/abs/2010Sci...330..944R} {330, 944}

\bibitem[\protect\citeauthoryear{{Rea} et~al.,}{{Rea} et~al.}{2013}]{rip+13}
{Rea} N.,  et~al., 2013, \mn@doi [\apj] {10.1088/0004-637X/770/1/65}, \href
  {https://ui.adsabs.harvard.edu/abs/2013ApJ...770...65R} {770, 65}

\bibitem[\protect\citeauthoryear{{Rigoselli}, {Mereghetti}, {Suleimanov},
  {Potekhin}, {Turolla}, {Taverna}  \& {Pintore}}{{Rigoselli}
  et~al.}{2019}]{rms+19}
{Rigoselli} M.,  {Mereghetti} S.,  {Suleimanov} V.,  {Potekhin} A.~Y.,
  {Turolla} R.,  {Taverna} R.,   {Pintore} F.,  2019, \mn@doi [\aap]
  {10.1051/0004-6361/201935485}, \href
  {https://ui.adsabs.harvard.edu/abs/2019A&A...627A..69R} {627, A69}

\bibitem[\protect\citeauthoryear{{Staveley-Smith} et~al.,}{{Staveley-Smith}
  et~al.}{1996}]{ParkesMultibeamReceiver}
{Staveley-Smith} L.,  et~al., 1996, \pasa, \href
  {https://ui.adsabs.harvard.edu/abs/1996PASA...13..243S} {13, 243}

\bibitem[\protect\citeauthoryear{{Tan} et~al.,}{{Tan} et~al.}{2018}]{tbc+18}
{Tan} C.~M.,  et~al., 2018, \mn@doi [ApJ] {10.3847/1538-4357/aade88}, \href
  {https://ui.adsabs.harvard.edu/#abs/2018ApJ...866...54T} {866, 54}

\bibitem[\protect\citeauthoryear{{Tauris} \& {Konar}}{{Tauris} \&
  {Konar}}{2001}]{tk01}
{Tauris} T.~M.,  {Konar} S.,  2001, \mn@doi [\aap]
  {10.1051/0004-6361:20010988}, \href
  {https://ui.adsabs.harvard.edu/abs/2001A&A...376..543T} {376, 543}

\bibitem[\protect\citeauthoryear{{Tauris} \& {Manchester}}{{Tauris} \&
  {Manchester}}{1998}]{tm98}
{Tauris} T.~M.,  {Manchester} R.~N.,  1998, \mn@doi [\mnras]
  {10.1046/j.1365-8711.1998.01369.x}, \href
  {https://ui.adsabs.harvard.edu/abs/1998MNRAS.298..625T} {298, 625}

\bibitem[\protect\citeauthoryear{{Thompson} \& {Duncan}}{{Thompson} \&
  {Duncan}}{1995}]{td95}
{Thompson} C.,  {Duncan} R.~C.,  1995, \mn@doi [\mnras]
  {10.1093/mnras/275.2.255}, \href
  {https://ui.adsabs.harvard.edu/abs/1995MNRAS.275..255T} {275, 255}

\bibitem[\protect\citeauthoryear{{Turolla}}{{Turolla}}{2009}]{Turolla2009}
{Turolla} R.,  2009, in {Becker} W.,  ed.,  Vol. 357, Astrophysics and Space
  Science Library. p.~141, \mn@doi{10.1007/978-3-540-76965-1_7}

\bibitem[\protect\citeauthoryear{{Vigan{\`o}}, {Rea}, {Pons}, {Perna},
  {Aguilera}  \& {Miralles}}{{Vigan{\`o}} et~al.}{2013}]{Vigano+2013}
{Vigan{\`o}} D.,  {Rea} N.,  {Pons} J.~A.,  {Perna} R.,  {Aguilera} D.~N.,
  {Miralles} J.~A.,  2013, \mn@doi [\mnras] {10.1093/mnras/stt1008}, \href
  {http://adsabs.harvard.edu/abs/2013MNRAS.434..123V} {434, 123}

\bibitem[\protect\citeauthoryear{{Weltevrede}}{{Weltevrede}}{2016}]{PSRSALSA}
{Weltevrede} P.,  2016, \mn@doi [\aap] {10.1051/0004-6361/201527950}, \href
  {https://ui.adsabs.harvard.edu/abs/2016A&A...590A.109W} {590, A109}

\bibitem[\protect\citeauthoryear{{Yao}, {Manchester}  \& {Wang}}{{Yao}
  et~al.}{2017}]{ymw16}
{Yao} J.~M.,  {Manchester} R.~N.,   {Wang} N.,  2017, \mn@doi [\apj]
  {10.3847/1538-4357/835/1/29}, \href
  {https://ui.adsabs.harvard.edu/abs/2017ApJ...835...29Y} {835, 29}

\bibitem[\protect\citeauthoryear{{Young}, {Manchester}  \& {Johnston}}{{Young}
  et~al.}{1999}]{ymj99}
{Young} M.~D.,  {Manchester} R.~N.,   {Johnston} S.,  1999, \mn@doi [\nat]
  {10.1038/23650}, \href
  {https://ui.adsabs.harvard.edu/abs/1999Natur.400..848Y} {400, 848}

\bibitem[\protect\citeauthoryear{{Young}, {Weltevrede}, {Stappers}, {Lyne}  \&
  {Kramer}}{{Young} et~al.}{2014}]{yws+14}
{Young} N.~J.,  {Weltevrede} P.,  {Stappers} B.~W.,  {Lyne} A.~G.,   {Kramer}
  M.,  2014, \mn@doi [MNRAS] {10.1093/mnras/stu1036}, \href
  {https://ui.adsabs.harvard.edu/#abs/2014MNRAS.442.2519Y} {442, 2519}

\bibitem[\protect\citeauthoryear{{Zhang}, {Harding}  \& {Muslimov}}{{Zhang}
  et~al.}{2000}]{zhm00}
{Zhang} B.,  {Harding} A.~K.,   {Muslimov} A.~G.,  2000, \mn@doi [\apjl]
  {10.1086/312542}, \href
  {https://ui.adsabs.harvard.edu/abs/2000ApJ...531L.135Z} {531, L135}

\bibitem[\protect\citeauthoryear{{van Straten} \& {Bailes}}{{van Straten} \&
  {Bailes}}{2011}]{vanStraten2011}
{van Straten} W.,  {Bailes} M.,  2011, \mn@doi [\pasa] {10.1071/AS10021}, \href
  {https://ui.adsabs.harvard.edu/abs/2011PASA...28....1V} {28, 1}

\makeatother
\end{thebibliography}







\bsp	
\label{lastpage}
\end{document}